%% file: sample-manuscript.tex
\documentclass[manuscript,screen]{acmart}

\usepackage{booktabs}
\usepackage{multirow}
\usepackage{enumitem}
\usepackage{subfigure}
\usepackage[ruled]{algorithm2e}
\usepackage[flushleft]{threeparttable}
\usepackage{bbm}
\usepackage{mathrsfs}
\usepackage{longtable}
\usepackage{lscape} 

\AtBeginDocument{%
  \providecommand\BibTeX{{%
    \normalfont B\kern-0.5em{\scshape i\kern-0.25em b}\kern-0.8em\TeX}}}

\usepackage{xcolor}
\usepackage{soul}
\usepackage{colortbl}

\newcommand{\eg}{\emph{e.g.}}
\newcommand{\ie}{\emph{i.e.}}

\definecolor{FeatEng}{rgb}{0.949,0.504,0.504}
\definecolor{FeatEco}{rgb}{0.582,0.879,0.824}
\definecolor{ScoreRank}{rgb}{0.836,0.965,0.676}
\definecolor{UserInter}{rgb}{0.996078,0.890196,0.54117}

\setcopyright{acmlicensed}
\copyrightyear{2018}
\acmYear{2018}
\acmDOI{XXXXXXX.XXXXXXX}

\acmConference[Conference acronym 'XX]{Make sure to enter the correct
  conference title from your rights confirmation emai}{June 03--05,
  2018}{Woodstock, NY}
\acmISBN{978-1-4503-XXXX-X/18/06}

\begin{document}

\title{How Can Recommender Systems Benefit from Large Language Models: A Survey}

\author{Jianghao Lin$^{*}$}
\affiliation{%
\institution{Shanghai Jiao Tong University}
\country{China}}
\email{chiangel@sjtu.edu.cn}

\author{Xinyi Dai$^{*}$}
\affiliation{%
\institution{Noah's Ark Lab, Huawei}
\country{China}}
\email{daixinyi3@huawei.com}

\author{Yunjia Xi}
\affiliation{%
\institution{Shanghai Jiao Tong University}
\country{China}}
\email{xiyunjia@sjtu.edu.cn}

\author{Weiwen Liu}
\author{Bo Chen}
\affiliation{%
	\institution{Noah's Ark Lab, Huawei}
	\country{China}}
\email{liuweiwen8@huawei.com}
\email{chenbo116@huawei.com}

\author{Hao Zhang}
\author{Yong Liu}
\affiliation{%
	\institution{Noah's Ark Lab, Huawei}
	\country{Singapore}}
\email{zhang.hao3@huawei.com}
\email{liu.yong6@huawei.com}

\author{Chuhan Wu}
\author{Xiangyang Li}
\affiliation{%
	\institution{Noah's Ark Lab, Huawei}
	\country{China}}
\email{wuchuhan1@huawei.com}
\email{lixiangyang34@huawei.com}

\author{Chenxu Zhu}
\author{Huifeng Guo}
\affiliation{%
	\institution{Noah's Ark Lab, Huawei}
	\country{China}}
\email{zhuchenxu1@huawei.com}
\email{huifeng.guo@huawei.com}

\author{Yong Yu}
\affiliation{%
  \institution{Shanghai Jiao Tong University}
  \country{China}}
\email{yyu@sjtu.edu.cn}

\author{Ruiming Tang$^\dagger$}
\affiliation{%
\institution{Noah's Ark Lab, Huawei}
\country{China}}
\email{tangruiming@huawei.com}

\author{Weinan Zhang$^\dagger$}
\affiliation{%
  \institution{Shanghai Jiao Tong University}
  \country{China}}
\email{wnzhang@sjtu.edu.cn}

\renewcommand{\shortauthors}{J. Lin et al.}

\begin{abstract}
With the rapid development of online services and web applications, recommender systems (RS) have become increasingly indispensable for mitigating information overload and matching users' information needs by providing personalized suggestions over items.
Although the RS research community has made remarkable progress over the past decades, conventional recommendation models (CRM) still have some limitations, \eg, lacking open-domain world knowledge, and difficulties in comprehending users' underlying preferences and motivations.
Meanwhile, large language models (LLM) have shown impressive general intelligence and human-like capabilities for various natural language processing (NLP) tasks, which mainly stem from their extensive open-world knowledge, logical and commonsense reasoning abilities, as well as their comprehension of human culture and society.
Consequently, the emergence of LLM is inspiring the design of recommender systems and pointing out a promising research direction, \ie, whether we can incorporate LLM and benefit from their common knowledge and capabilities to compensate for the limitations of CRM.
In this paper, we conduct a comprehensive survey on this research direction, and draw a bird's-eye view from the perspective of the whole pipeline in real-world recommender systems.
Specifically, we summarize existing research works from two orthogonal aspects: where and how to adapt LLM to RS.
For the ``\textbf{WHERE}'' question, we discuss the roles that LLM could play in different stages of the recommendation pipeline, \ie, feature engineering, feature encoder, scoring/ranking function, user interaction, and pipeline controller. 
For the ``\textbf{HOW}'' question, we investigate the training and inference strategies, resulting in two fine-grained taxonomy criteria, \ie, whether to tune LLM or not during training, and whether to involve conventional recommendation models for inference.
Detailed analysis and general development paths are provided for both ``WHERE'' and ``HOW'' questions, respectively.
Then, we highlight the key challenges in adapting LLM to RS from three aspects, \ie, efficiency, effectiveness, and ethics. 
Finally, we summarize the survey and discuss the future prospects.
To further facilitate the research community of LLM-enhanced recommender systems, we actively maintain a GitHub repository for papers and other related resources in this rising direction\footnote{\url{https://github.com/CHIANGEL/Awesome-LLM-for-RecSys}}.
\let\thefootnote\relax\footnotetext{* Jianghao Lin and Xinyi Dai are the co-first authors.}
\let\thefootnote\relax\footnotetext{$\dagger$ Ruiming Tang and Weinan Zhang are the co-corresponding authors.}

\end{abstract}



\received{20 February 2007}
\received[revised]{12 March 2009}
\received[accepted]{5 June 2009}

\maketitle

\input{sections/introduction_new_new}

\input{sections/preliminary}

\input{sections/where-to-adapt-LLMs}

\input{sections/how-to-adapt-LLMs}

\input{sections/challenge}

\input{sections/conclusion}

\begin{acks}
The Shanghai Jiao Tong University team is partially supported by Shanghai Municipal Science and Technology Major Project (2021SHZDZX0102) and National Natural Science Foundation of China (62322603, 62177033).
Jianghao Lin is supported by the  Wu Wen Jun Honorary Doctoral Scholarship. 
The work is sponsored by Huawei Innovation Research Program.
We thank MindSpore~\cite{mindspore} for the partial support of this work, which is a new deep learning computing framework.
\end{acks}

\bibliographystyle{ACM-Reference-Format}
\bibliography{sample-base}

\input{sections/appendix}

\end{document}

%% file: sections/introduction_new_new.tex
\section{Introduction}

With the rapid development of online services, recommender systems (RS) have become increasingly important to match users' information needs~\cite{dai2021adversarial,fu2023f,liu2024mamba4rec} and mitigate information overload~\cite{guo2017deepfm,lin2023map}. 
They offer personalized suggestions across diverse domains such as e-commerce~\cite{schafer2001commerce}, movie~\cite{goyani2020review}, music~\cite{song2012survey}, etc.
Despite the varied forms of recommendation tasks (\eg, top-$N$ recommendation, and sequential recommendation), the common learning objective for recommender systems is to estimate a given user's preference towards each candidate item, and finally arrange a ranked list of items presented to the user~\cite{lin2021graph,xi2023bird}. 

Despite the remarkable progress of conventional recommender systems over the past decades, their recommendation performance is still suboptimal, hampered by two major drawbacks as follows: 
(1) Conventional recommender systems are domain-oriented systems generally built based on discrete ID features within specific domains~\cite{xi2023towards}. 
Therefore, they lack open-domain world knowledge to obtain better recommendation performance (\eg, enhancing user interest modeling and item content understanding), and transferring abilities across different domains and platforms~\cite{harte2023leveraging,liu2023first,chen2023knowledge}. 
(2) Conventional recommender systems often aim to optimize specific user feedback such as clicks and purchases in a data-driven manner, where the user preference and underlying motivations are often implicitly modeled based on user behaviors collected online. As a result, these systems might lack recommendation explainability~\cite{gao2023chat,chen2023survey}, and cannot fully understand the complicated and volatile intent of users in various contexts. 
Moreover, users cannot actively guide the recommender system to follow their requirements and customize recommendation results by providing detailed instructions in natural language~\cite{friedman2023leveraging,wang2022recindial,wang2022barcor}.

With the emergence of large foundation models in recent years, they provide promising and universal insights when handling many challenging problems in the data mining field~\cite{chen2023large,tan2023user,fu2023codeapex}.
A representative form is the large language model (LLM), which has shown impressive general intelligence in various language processing tasks due to their huge memory of open-world knowledge, the ability of logical and commonsense reasoning, and the awareness of human society and culture~\cite{zhao2023survey, huang2022towards, brown2020language}.
By using natural language as a universal information carrier, knowledge in different forms, modalities, domains, and platforms can be generally integrated, exploited, and interpreted.
Consequently, the rise of large language models is inspiring the design of recommender systems, \ie, whether we can incorporate LLM and benefit from their common knowledge to address the aforementioned ingrained drawbacks of conventional recommender systems. 



\begin{figure}[t]
    \centering
    \includegraphics[width=0.99\textwidth]{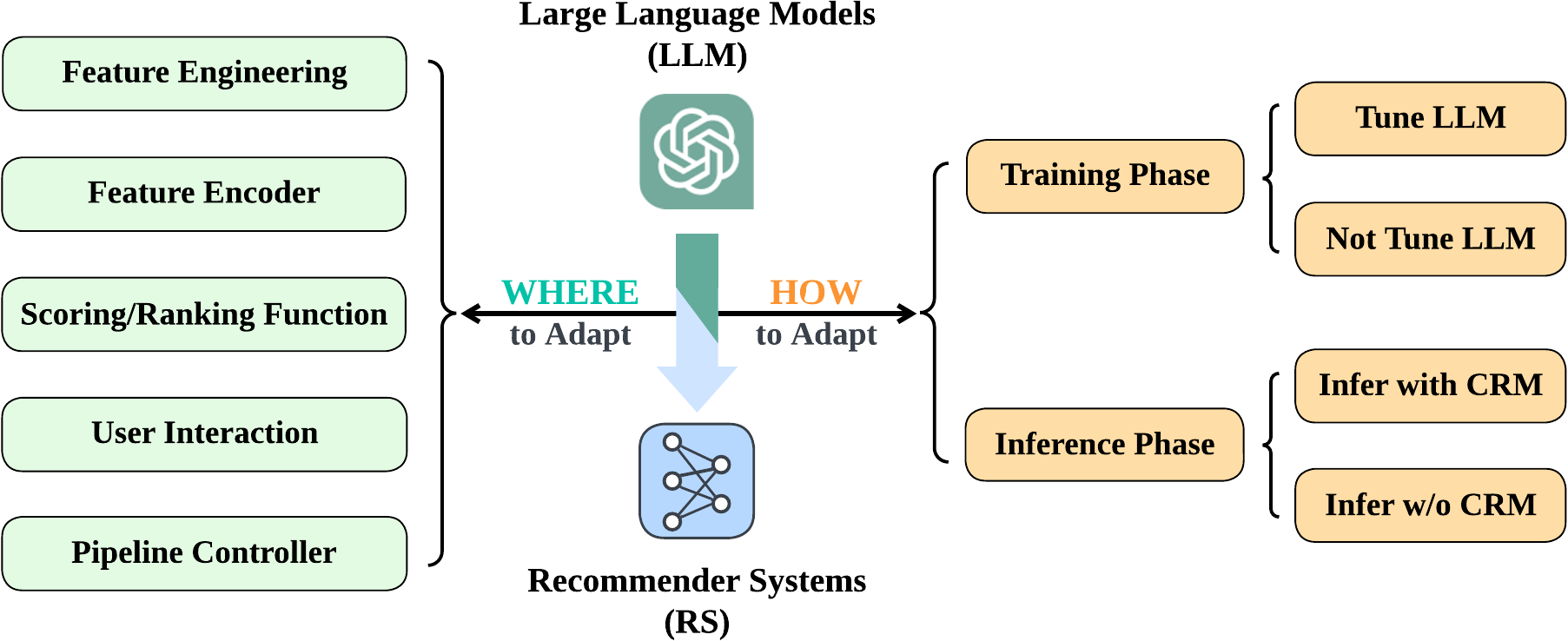}
    \caption{The decomposition of our core research question about adapting large language models to recommender systems. We analyze the question from two orthogonal perspectives: (1) where to adapt LLM, and (2) how to adapt LLM. Note that CRM stands for conventional recommendation model.}
    \label{fig:overall two perspectives}
\end{figure}

Recently, RS researchers and practitioners have made many pioneer attempts to employ LLM in current recommendation pipelines, and have achieved notable progress in boosting the performance of different canonical recommendation processes such as feature modeling~\cite{xi2023towards} and ranking~\cite{bao2023tallrec}.
There exist several related survey works that delve into the potential of LLM for general recommender systems.
\citet{wu2023survey} conduct a review on both discriminative and generative LLMs for recommendation with different tuning strategies.
\citet{fan2023recommender} focus on the pretraining, finetuning and prompting approaches when leveraging LLM for recommendation. 
\citet{huang2024foundation} investigate the recommendation foundation models from aspects of both different model types and various downstream tasks. 
Other works also concentrate on one specific aspect of recommender systems with LLM enhancements, \eg, prompting strategy~\cite{xu2024prompting}, generative recommendation~\cite{li2024survey,li2023large,deldjoo2024review}, and explainable recommendation~\cite{chen2023survey}.
However, it still lacks a bird's-eye view of where and how recommender systems can embrace large language models and integrate them into the overall recommendation pipeline, which is essential in building a technique map to systematically guide the research, practice, and service in LLM-empowered recommendation.

Different from existing surveys on this topic, in this paper, we propose a systematic view of the LLM-enhanced recommendation, from the angle of the whole pipeline in industrial recommender systems. LLM is currently utilized in various stages of recommendation systems and are integrated with current systems via different techniques.
To conduct a comprehensive review of latest research progress, as shown in Figure~\ref{fig:overall two perspectives}, we propose research questions about LLM-enhanced recommender systems from the following two perspectives:

\begin{itemize}[leftmargin=10pt]
    \item \textbf{``WHERE''} question focuses on where to adapt LLM for RS, and discusses the roles that LLM could play at different stages of current recommender system pipeline, \ie, feature engineering, feature encoder, scoring/ranking function, user interaction, and pipeline controller. 
    \item \textbf{``HOW''} question centers on how to adapt LLM for RS, where two orthogonal taxonomy criteria are carried out: (1) whether we will freeze the parameters of the large language model during the training phase, and (2) whether we will involve conventional recommendation models (CRM) during the inference phase.
\end{itemize}

From the two perspectives, we propose feasible and instructive suggestions for the evolution of existing online recommendation platforms in the era of large language models\footnote{To provide a thorough survey and a clear development path, we broaden the scope of large language models, and bring those relatively smaller language models (\eg, BERT~\cite{devlin2018bert}, GPT2~\cite{radford2019language}) into the discussion as well.}\footnote{ We focus on works that leverage LLM together with their pretrained parameters to handle textual features via prompting, and exclude works that simply apply pretraining paradigms from NLP domains to pure ID-based traditional recommendation models (\eg, BERT4Rec~\cite{sun2019bert4rec}). Interested readers can refer to \cite{yu2022self,liu2023pre}.}.

The rest of this paper is organized as follows. 
In Section~\ref{sec:preliminary}, we briefly introduce the background and preliminary for recommender systems and large language models.
Section~\ref{sec: where} and Section~\ref{sec: how} thoroughly analyze the aforementioned taxonomies from two perspectives (\ie, \textbf{``WHERE''} and \textbf{``HOW''}), followed by detailed discussion and analysis of the general development path.
In Section~\ref{sec: challenge}, we highlight the key challenges and future directions for the adaption of LLM to RS from three aspects (\ie, \textbf{efficiency}, \textbf{effectiveness}, and \textbf{ethics}), which mainly arise from the real-world applications of recommender systems.
Finally, Section~\ref{sec:conclusion} concludes this survey and draws a hopeful vision for future prospects in research communities of LLM-enhanced recommender systems. 
Furthermore, we give a comprehensive look-up table of related works that adapt LLM to RS in Appendix~\ref{sec:look up table} (\ie, Table~\ref{tab:look up table}), attaching the detailed information for each work, \eg, the stage that LLM is involved in, LLM backbone, and LLM tuning strategy, etc.

%% file: sections/preliminary.tex
\section{Background and Preliminary}
\label{sec:preliminary}

Before elaborating on the detail of our survey, we would like to introduce the following background and basic concepts: (1) the general pipeline of modern recommender systems based on deep learning techniques, and (2) the general workflow and concepts for large language models. 

\subsection{Modern Recommender Systems}
\label{sec:preliminary rs}

The core task of recommender systems is to provide a ranked list of items $[i_k]_{k=1}^N, i_k \in \mathcal{I}$ for the user $u\in\mathcal{U}$ given a certain context $c$, where $\mathcal{I}$ and $\mathcal{U}$ are the universal sets of items and users, respectively. 
Note that scenarios like next item prediction are special cases for such a formulation with $N=1$.
We denote the goal as follows:
\begin{equation}
    [i_k]_{k=1}^N\leftarrow \operatorname{RS}(u,c,\mathcal{I}),\; u\in\mathcal{U},i_k\in\mathcal{I}.
\end{equation}

\begin{figure}[t]
    \centering
    \includegraphics[width=0.99\textwidth]{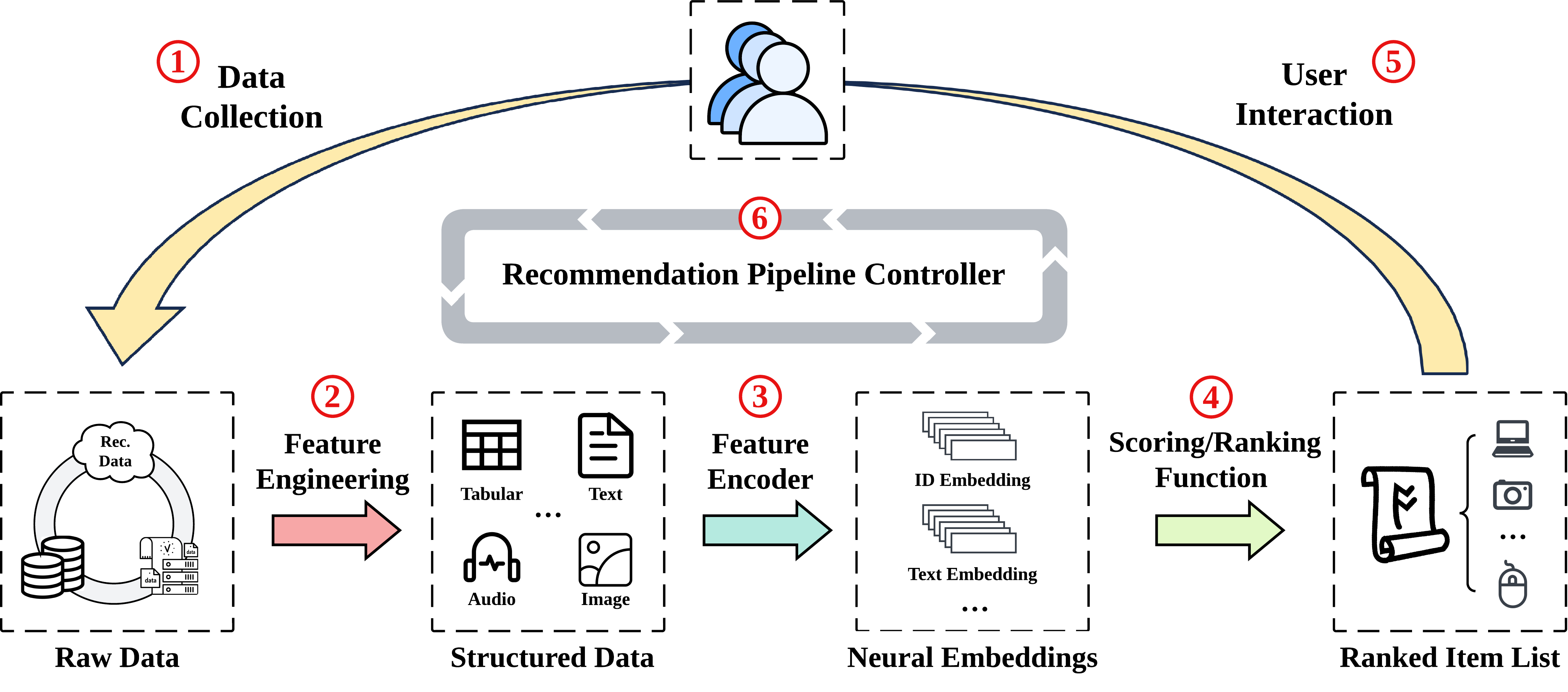}
    \caption{The illustration of deep learning based recommender system pipeline. 
    We characterize the modern recommender system as an information cycle that consists of six stages: data collection, feature engineering, feature encoder, scoring/ranking function, user interaction, and recommendation pipeline controller, which are denoted by different colors.}
    \label{fig:modern rs illustration}
\end{figure}

As shown in Figure~\ref{fig:modern rs illustration}, the modern deep learning based recommender systems can be characterized as an information cycle that encompasses six key stages: (1) Data Collection, where the users' feedback data is gathered; (2) Feature Engineering, which involves preparing and processing the collected raw data; (3) Feature Encoder, where data features are transformed into neural embeddings; (4) Scoring/Ranking Function, which selects and orders the recommended items; (5) User Interaction, which determines how users engage with the recommendations; and finally, (6) Recommendation Pipeline Controller, which serves as the central mechanism tying all the stages above together in a cohesive process. 
Next, we will briefly go through each of the stages as follows: \

\begin{itemize}[leftmargin=10pt]
    \item \textbf{Data Collection.} The data collection stage gathers both explicit and implicit feedback from online services by presenting recommended items to users. 
    The explicit feedback indicates direct user responses such as ratings, while the implicit feedback is derived from user behaviors like clicks, downloads, and purchases. 
    In addition to gathering user feedback, the data to be collected also encompasses a range of raw features including item attributes, user demographics, and contextual information. 
    The raw data is stored in the database in certain formats such as JSON for further processing. 
    It is worth noting that, in this paper, the data collection mainly refers to the process of collecting real-world user behavior data from online services \textbf{without} any manual transformation or data synthesis. 
    \item \textbf{Feature Engineering.} Feature engineering is the process of selecting, manipulating, transforming, and augmenting the raw data collected online into structured data that is suitable as inputs of neural recommendation models. 
    As shown in Figure~\ref{fig:modern rs illustration}, the major outputs of feature engineering consist of various forms of features, which will be then encoded by feature encoders of different modalities, \eg, language models for textual features, vision models for visual features, and conventional recommendation models (CRM) for ID features. 
    Note that the feature engineering stage stands for the general data engineering process, which does not only conduct feature-level manipulation, but also involves sample-level synthesizing and augmentation for enhancements of both training and offline evaluation.
    \item \textbf{Feature Encoder.}
    Generally speaking, the feature encoder takes as input the processed features from the feature engineering stage, and generates the corresponding neural embeddings for scoring/ranking functions in the next stage. 
    Various encoders are employed depending on the data modality. Typically, this process is executed as an embedding layer for one-hot encoded categorical features in standard recommendation models. Features of other modalities, such as text, vision, video, or audio, are further used and encoded to enhance content understanding.
    \item \textbf{Scoring/Ranking function.} Scoring/Ranking function serves as the core part of recommendation to select or rank the top-relevant items to satisfy users' information needs based on the neural embeddings generated by the feature encoders. Researchers develop various neural methods to precisely estimate the user preference and behavior patterns based on various techniques, \eg, collaborative filtering~\cite{su2009survey,he2017neural}, sequential modeling~\cite{cheng2024general,liu2023user}, graph neural networks~\cite{wang2018ripplenet,wang2019neural}, etc.
    \item \textbf{User Interaction.}
    User interaction refers to the way we represent the recommended items to the target user, and the way users give their feedback back to the recommender system. 
    While traditional recommendation pages basically involve a single list of items, various complex and multi-modal scenarios are recently proposed and studied~\cite{zaafira2023siak}. 
    For example, conversational recommendation provides natural language interface and enables multi-round interactive recommendation for the user~\cite{sun2018conversational}.
    Besides, multi-block page-level user interactions are also widely considered for nested user feedback~\cite{fu2023f,sagtani2024learning}.
    \item \textbf{Recommendation Pipeline Control.} Pipeline controller monitors and controls the operations of the whole recommendation pipeline mentioned above. It can even provide fine-grained control over different stages for recommendation (\eg, matching, ranking, reranking), or decide to combine different downstream models and APIs to accomplish the final recommendation tasks.
\end{itemize}

It is worth noting that the multi-stage formulation above serves as a general overview of the modern recommendation pipeline, and some of the stages might be skipped, linked, or merged with specific modeling techniques. For example, if we adapt LLM as the scoring/ranking function, the input of the recommender should be textual data from the feature engineering~\cite{lin2023rella,bao2023tallrec}, instead of neural embeddings from the feature encoder, \ie, we skip the feature encoder stage. 
Moreover, some works~\cite{zhang2023collm,li2023e4srec} would explore advanced techniques to inject domain knowledge of neural embeddings from CRM into LLM as the scoring/ranking function, \ie, we obtains the inputs by merging the outcomes from both feature engineering and feature encoder stages.

\subsection{Large Language Models}

Language models aim to conduct the probabilistic modeling of natural languages to predict the word tokens given a specific textual context. 
Nowadays, most language models are built based on transformer-like~\cite{vaswani2017attention} architectures to proficiently model the context dependency for human languages. 
They are first pretrained on a massive amount of unlabeled text data, and then further finetuned with task-oriented data for different downstream applications.
These pretrained language models (PLM) can be mainly classified into three categories: encoder-only models like BERT~\cite{devlin2018bert}, decoder-only models like GPT~\cite{radford2019language}, and encoder-decoder models like T5~\cite{raffel2020exploring}.

Large language models (LLM) are the scaled-up derivatives of traditional pretrained language models mentioned above, in terms of both model sizes and data volumes, \eg, GPT-3~\cite{brown2020language}, PaLM~\cite{chowdhery2023palm}, LLaMA~\cite{touvron2023llama}, ChatGLM~\cite{du2022glm,zeng2022glm}. 
A typical LLM usually consists of  billion-level or even trillion-level parameters, and is pretrained on much larger volumes of textual corpora with up-to trillions of tokens crawled from various Internet sources like Wikipedia, GitHub, ArXiv, etc. 
As illustrated by the scaling law~\cite{kaplan2020scaling,hoffmann2022training}, the scaling up of model size, data volume and training scale can continuously contribute to the growth of model performance for a wide range of downstream NLP tasks. 
Furthermore, researchers find that LLM can exhibit emergent abilities, \eg, few-shot in-context learning, instruction following and step-by-step reasoning, when the model size continues to scale up and reaches a certain threshold~\cite{wei2022emergent}

LLM has revolutionized the field of NLP by demonstrating impressive capabilities in understanding natural languages and generating human-like texts. 
Moreover, LLM has gone beyond the field of NLP and shown remarkable potential in various deep learning based applications, such information system~\cite{zhu2024large}, education~\cite{li2023adapting}, finance~\cite{wu2023bloomberggpt} and healthcare~\cite{nov2023putting,tang2023does}. 
Therefore, recent studies start to investigate the application of LLM to recommender systems. 
Equipped with the extensive open-world knowledge and powerful emergent abilities like reasoning, LLM is able to analyze the individual preference based on user behavior sequences, and promote the content understanding and expansion for items, which can largely enhance the recommendation performance~\cite{xi2023towards,bao2023tallrec,cui2022m6,yang2023large}.
Besides, LLM can also support more complex scenarios like conversational recommendation~\cite{gao2023chat}, explainable recommendation~\cite{chen2023survey}, as well as task decomposition and tool usage (\eg, search engines)~\cite{wang2023recmind} for recommendation enhancements.

%% file: sections/where-to-adapt-LLMs.tex
\section{Where to Adapt Large Language Models}
\label{sec: where}

Based on the decomposition of modern recommender systems discussed in Section~\ref{sec:preliminary rs}, we answer the \textbf{``WHERE''} question by elaborating on the adaptation of LLM to different parts of the recommendation pipeline: (1) feature engineering, (2) feature encoder, (3) scoring/ranking function, (4) user interaction, and (5) pipeline controller. 
It is worth noting that, the utilization of LLM in the same research work may involve multiple stages of the recommendation pipeline due the multi-task nature of LLM. For example, LLM is leveraged in both stages of feature engineering and scoring/ranking function in CUP~\cite{torbati2023recommendations}.



\begin{figure}[t]
    \centering
    \includegraphics[width=0.99\textwidth]{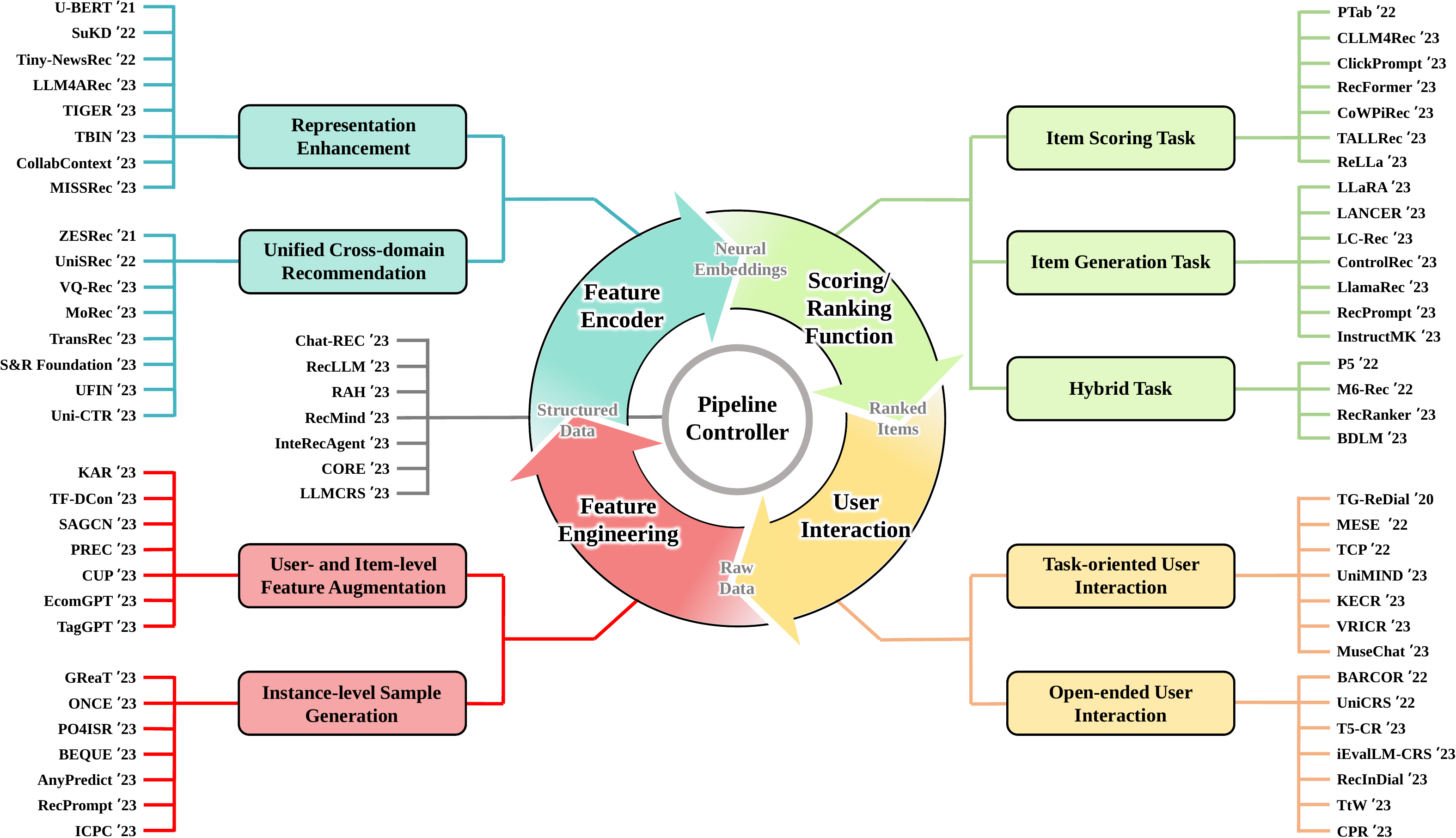}
    \caption{
    The illustrative dissection of the ``\textbf{WHERE}'' research question. 
    We show that LLM can be adapted to different stages of the recommender system pipeline as introduced in Section~\ref{sec:preliminary rs}, \ie, feature engineering, feature encoder, scoring/ranking function, user interaction, and pipeline controller.
    We provide finer-grained classification criteria for each stage, and list representative works denoted by different colors.}
    \label{fig:where LLM RS}
\end{figure}

\subsection{LLM for Feature Engineering}

In the feature engineering stage, LLM takes as inputs the original features (\eg, item descriptions, user profiles, and user behaviors), and generates auxiliary textual features for data augmentation with varied goals, \eg, enriching the training data, alleviating the long-tail problem, etc.
Different prompting strategies are employed to make full use of the open-world knowledge and reasoning ability exhibited by LLM. 
According to the type of data augmentation, the research works of this line can be mainly classified into two categories: (1) user- and item-level feature augmentation, (2) instance-level training sample generation. 

\subsubsection{User- and Item-level Feature Augmentation}

Equipped with powerful reasoning ability and open-world knowledge, LLM is often treated as a flexible knowledge base~\cite{luo2024integrating}. Hence, it can provide auxiliary features for better user preference modeling and item content understanding. 
As a representative, KAR~\cite{xi2023towards} adopts LLM to generate the user-side preference knowledge and item-side factual knowledge, which serve as the plug-in features for downstream conventional recommendation models.
SAGCN~\cite{liu2023understanding} introduces a chain-based prompting approach to uncover semantic aspect-aware interactions, which provides clearer insights into user behaviors at a fine-grained semantic level.
CUP~\cite{torbati2023recommendations} adopts ChatGPT to summarize each user's interests with a few short keywords according to the user review texts. In this way, the user profiling data is condensed within 128 tokens and thus can be further encoded with small-scale language models that are constrained by the context windows size (\eg, 512 for BERT~\cite{devlin2018bert}).
Moreover, instead of using a frozen LLM for feature augmentation, LLaMA-E\cite{shi2023llama} and EcomGPT~\cite{li2023ecomgpt} finetune the base large language models for various downstream generative tasks in e-commerce scenarios, \eg, product categorization and intent speculation.
Other works also utilize LLM to further enrich the training data from different perspectives, \eg, text refinement~\cite{du2023enhancing,zheng2023generative,liu2023modeling}, knowledge graph completion and reasoning~\cite{chen2023knowledge,wang2023enhancing,wei2023llmrec,chu2024llm}, attribute generation~\cite{li2023taggpt,brinkmann2023product,yin2023heterogeneous}, and user interest modeling~\cite{christakopoulou2023large,lyu2023llm,ren2023representation,doddapaneni2024user}. 

\subsubsection{Instance-level Sample Generation}

Apart from feature-level augmentations, LLM is also leveraged to generate synthetic samples, which enrich the training dataset~\cite{mysore2023large} and improve the model prediction quality~\cite{sun2023large,liu2023recprompt}. 
GReaT~\cite{borisov2023language} tunes a generative language model to synthesize realistic tabular data as augmentations for the training phase.
\citet{carranza2023privacy} explore to train a differentially private (DP) large language model for synthetic user query generation, in order to address the privacy problem in recommender systems.
ONCE~\cite{liu2023first} applies manually designed prompts to obtain additional news summarization, user profiles, and synthetic news pieces for news recommendation.
AnyPredict~\cite{wang2023anypredict} leverages LLM to consolidate datasets with different feature fields, and align out-domain datasets for a shared target task. 
TF-DCon~\cite{wu2023leveraging} aims to compress and condensate the training data by using LLM to generate fewer amount of synthetic samples from views of both user history and item content.
\citet{zhang2023generative} further attempt to incorporate multiple large language models as agents to simulate the fine-grained user communication and interaction for more realistic recommendation scenarios. 
Moreover, RecPrompt~\cite{liu2023recprompt} and PO4ISR~\cite{sun2023large} propose to perform automatic prompt template optimization with powerful LLM (\eg, ChatGPT or GPT4), and therefore iteratively improve the recommendation performance with gradually better textual inputs for LLM-based recommenders. 
BEQUE~\cite{peng2023large} finetunes and deploys LLM for query rewriting in e-commercial scenarios to bridge the semantic gaps inherent in the semantic matching process, especially for long-tail queries.
\citet{li2023agent4ranking} use Chain-of-Thought~\cite{wei2022chain}
(CoT) technology to leverage LLM as
agent to emulate various demographic profiles for robust and efficient query rewriting.

\subsection{LLM as Feature Encoder}

In conventional recommender systems, the structured data are usually formulated as one-hot encodings, and a embedding layer is adopted as the feature encoder to obtain dense embeddings. 
With the emergence of language models, researchers propose to adopt LLM as auxiliary textual feature encoder to gain two major benefits: (1) further enriching the user/item representations with semantic information for the later neural recommendation models; (2) achieving cross-domain\footnote{Different domains means data sources with different distributions, \eg, scenarios, datasets, platforms, etc.} recommendation with natural language as the bridge, where ID feature fields might not be shared.

\subsubsection{Representation Enhancement}

For item representation enhancement, LLM is leveraged as feature encoder for scenarios with abundant textual features available (\eg, item title, body text, detailed description), including but not limited to: document ranking~\cite{zou2021pre,liu2021pre}, news recommendation~\cite{wu2021empowering,wu2022mm,yu2022tiny,liu2022boosting,runfeng2023lkpnr}, tweet search~\cite{zhang2022twhin}, tag selection~\cite{he2022ptm4tag}, For 
While the item content is generally static, the user interest is highly dynamic and keeps evolving over time, therefore requiring sequential modeling over the fast-evolving user behaviors and underlying preferences~\cite{pi2020search,zhou2018deep,kang2018self,du2024disco}.
For example, U-BERT~\cite{qiu2021u} ameliorates the user representation by encoding review texts into a sequence of dense vectors via BERT~\cite{devlin2018bert}, followed by specially designed attention networks for user interest modeling.
LLM4ARec~\cite{li2023prompt} uses GPT2~\cite{radford2019language} to extract personalized aspect terms and latent vectors from user profiles and reviews to better assist recommendations. 
In some special cases, the semantic representation encoded by LLM is not directly used as the input for the later scoring/ranking function. Instead, it is converted into a sequence of discrete tokens through quantization to adapt to scoring/ranking functions that require discrete inputs (\eg, generative recommendation). 
TIGER~\cite{rajput2023recommender} proposes to apply vector quantization techniques~\cite{zeghidour2021soundstream,yu2021vector,oord2018neural} over the semantic item representations to further compress each item into a tuple of discrete semantic tokens. 
Hence, the sequential recommendation can be expressed as a sequence modeling task over a list of discrete tokens, where classical transformer~\cite{vaswani2017attention} architectures can be employed.
Based on the idea of item vector quantization, LMIndexer~\cite{jin2023language} designs a self-supervised semantic indexing framework to capture the item's semantic representation and the corresponding semantic tokens at the same time in an end-to-end manner.

\subsubsection{Unified Cross-domain Recommendation}

Apart from the user/item representation improvement, adopting LLM as feature encoder also enables transfer learning and cross-domain recommendation, where natural language serves as the bridge to align the heterogeneous information from different domains~\cite{li2023multi,wang2022transrec,li2023tcf}.
ZESRec~\cite{ding2021zero} applies BERT~\cite{devlin2018bert} to convert item descriptions into universal semantic representations for zero-shot recommendation. 
In UniSRec~\cite{hou2022towards}, the item representations are learned for cross-domain sequential recommendation via a fixed BERT model followed by a lightweight MoE-enhanced network.
Built upon UniSRec, VQ-Rec~\cite{hou2023learning} introduces vector quantization techniques to better align the textual embeddings generated by LLM to the recommendation space.
Uni-CTR~\cite{fu2023unified} leverages layer-wise
semantic representations from a shared LLM to sufficiently capture the commonalities among different domains, which leads to better multi-domain recommendation.
Other works~\cite{tian2023ufin,gong2023unified} leverage unified cross-domain textual embeddings from a fixed LLM (\eg, ChatGLM~\cite{du2022glm}, Sheared-LLaMA~\cite{xia2023sheared}) to tackle scenarios with cold-start users/items or low-frequency long-tail features. 
\citet{fu2023exploring} further explore layerwise adapter tuning on large language models to obtain better embeddings over textual features from different domains.

\subsection{LLM as Scoring/Ranking Function}
\label{sec:scoring/ranking function}

The ultimate goal of the scoring/ranking stage is highly tied with the general purpose of recommender systems as discussed in Section~\ref{sec:preliminary rs}, \ie, to provide a ranked list of items $[i_k]_{k=1}^N, i_k \in \mathcal{I}$ for target user $u\in \mathcal{U}$, where $\mathcal{I}$ and $\mathcal{U}$ are the universal set of items and users (next item prediction is a special case where $N=1$). 
When directly adapting LLM as the scoring/ranking function, such a goal could be achieved through various kinds of tasks for LLM (\eg, rating prediction, item ID generation).
According to different tasks that LLM solves, we classify related research works into three categories: (1) item scoring task, (2) item generation task, and (3) hybrid task. 
Moreover, as discussed in Section~\ref{sec:preliminary rs}, when adapting LLM as the scoring/ranking function, the input for this stage can be textual data, neural embeddings from other encoders, or a combination of both. In this section, we mainly focus on the task formulations and solution paradigms of LLM for scoring \& ranking. We would omit the input format unless necessary.


\subsubsection{Item Scoring Task}

In item scoring tasks, the large language model serves as a pointwise function $F(u, i),\forall u \in \mathcal{U},\forall i \in \mathcal{I}$, which estimates the utility score of each candidate item $i$ for the target user $u$. Here $\mathcal{U}$ and $\mathcal{I}$ denote the universal set of users and items, respectively.
The final ranked list of items is obtained by sorting the utility score calculated between the target user $u$ and each item $i$ in the candidate set $\mathcal{C}$: 
\begin{equation}
\begin{aligned}
    \mathcal{C} &\leftarrow \operatorname{Pre-filter}(u,\mathcal{I}),\\
    [i_k]_{k=1}^N &\leftarrow \operatorname{Sort}\left(\{F(u,i)\;|\;\forall i\in\mathcal{I}\}\right),\;N\le |\mathcal{C}|,
\end{aligned}
\end{equation}
where $\mathcal{C}$ is the candidate set obtained via a pre-filter function (\eg, the retrieval and pre-ranking models for the ranking stage). 
The pre-filtering is conducted to reduce the number of candidate items, thus saving the computational cost.
The pre-filter can be an identity-mapping function (\ie, $\mathcal{C}=\mathcal{I}$) for the first retrieval stage for recommender systems.

Without loss of generality, the large language model takes as inputs the discrete tokens of textual prompt $x$, and generates the target token $\hat{t}$ as the output for either the masked token in masked language modeling or the next token in causal language modeling. 
The process can be formulated as follows:
\begin{equation}
\begin{aligned}
    h &= \operatorname{LLM}(x),\\
    s &= \operatorname{LM\_Head}(h)\in\mathbb{R}^V, \\
    p &= \operatorname{Softmax}(s)\in\mathbb{R}^V,\\
    \hat{t} &\sim p,
\end{aligned}
\label{eq:llm process}
\end{equation}
where $h$ is the final representation, $V$ is the vocabulary size, and $\hat{t}$ is the predicted token sampled from the probability distribution $p$.

However, the item scoring task requires the model to do pointwise scoring for a given user-item pair $(u,i)$, and the output should be a real number $\hat{y}=F(u,i)$, instead of generated discrete tokens $\hat{t}$.
The output $\hat{y}$ should fall within a certain numerical range to indicate the user preference, \eg, $\hat{y}\in[0,1]$ for click-through rate (CTR) estimation and $\hat{y}\in[0,5]$ for rating prediction. 
There are three major approaches to address such an issue that the output requires continuous numerical values while LLM produces discrete tokens.

The first type of solution~\cite{li2023e4srec,wang2023flip,lin2023clickprompt,wang2023bert4ctr,zhu2023collaborative,liu2022ptab,kang2023llms,lei2023recexplainer,zhuang2023setwise} adopts the single-tower paradigm~\cite{qu2016product,zhang2021deep}.
To be specific, they directly abandon the language modeling decoder head (\ie, $\operatorname{LM\_Head}(\cdot)$), and feed the final representation $h$ of LLM in Eq.~\ref{eq:llm process} into a delicately designed projection layer to calculate the final score $\hat{y}$ for classification or regression tasks, \ie, 
\begin{equation}
    \hat{y}=F(u,i)=\operatorname{MLP}(h),
\end{equation}
where MLP (short for multi-layer perceptron) is the projection layer. 
The input prompt $x$ needs to contain information from both the user $u$ and item $i$ to support the preference estimation based on one single latent representation $h$. 
E4SRec~\cite{li2023e4srec} constructs personalized prompts with the help of pre-learned user \& item ID embeddings for precise preference estimation.
FLIP~\cite{wang2023flip} and ClickPrompt~\cite{lin2023clickprompt} propose to conduct fine-grained knowledge alignment and fusion over the semantic and collaborative information in parallel and stacking paradigms, respectively.
CER~\cite{raczynski2023problem} reinforces the coherence between recommendations and their natural language explanations to improve the rating prediction performance.
\citet{kang2023llms} finetune the large language model for rating prediction in a regression manner, which exhibits a surprising performance by scaling the model size of finetuned LLM up to 11 billion.
Other typical examples in this line of research include:
LSAT~\cite{shi2023preliminary}, 
BERT4CTR~\cite{wang2023bert4ctr},
CLLM4Rec~\cite{zhu2023collaborative}, and
PTab~\cite{liu2022ptab}.

Similar to the first method, the second type of solution~\cite{torbati2023unveiling,liu2023text,tang2023one,torbati2023recommendations,yang2023collaborative,li2023text} also discards the decoder head of LLM. However, what sets it apart is that it adopts the popular two-tower structure~\cite{he2020lightgcn,wang2019neural,he2017neural} in conventional recommender systems.
They maintain both two separate towers to obtain the representations for user and item respectively, and the preference score is calculated via a certain distance metric between the two representations:
\begin{equation}
    \hat{y}=F(u,i)=d\left(T_u\left(x_{u}\right),T_i\left(x_{i}\right)\right),
\end{equation}
where $d(\cdot,\cdot)$ is the distance metric function (\eg, cosine similarity, L2 distance). 
$T_u(\cdot)$ and $T_i(\cdot)$ are the user and item towers that consist of LLM backbones to extract the useful knowledge representations from both user and item texts (\ie, $x_u$ and $x_i$).
In this line of works, different auxiliary structures are designed to augment the dual-side information with LLM.
For example, CoWPiRec~\cite{yang2023collaborative} applies word graph neural networks to item texts within the user behavior sequence to amplify the semantic information correlation.
By employing the encoder-decoder LLM, TASTE~\cite{liu2023text} first encodes each user behavior into a soft prompt vector and then leverages the decoder to extract the user preference from the sequence of soft prompts.
Other typical examples include: 
RecFormer~\cite{li2023text},
LLM-Rec~\cite{tang2023one}, and
CUP~\cite{torbati2023recommendations}. 

Different from the aforementioned two solutions that both replace the original language modeling decoder head (\ie, $\operatorname{LM\_Head}(\cdot)$) with manually designed predictive modules, the last type of solution~\cite{wang2024towards,chen2024elcorec,lin2024large,lin2023rella,wu2023towards,ho2023btrec,prakash2023cr,zhuang2023beyond,hegselmann2023tabllm,zhang2023prompt,wu2023exploring,sun2023instruction,qu2023thoroughly,mao2023unitrec,sileo2022zero,zhang2021language,zhiyuli2023bookgpt,luo2024integrating,xi2024play} proposes to preserve the decoder head and perform preference estimation based on the probability distribution $p\in\mathbb{R}^V$. 
TALLRec~\cite{bao2023tallrec}, ReLLa~\cite{lin2023rella}, CoLLM~\cite{zhang2023collm}, PromptRec~\cite{wu2023towards}, BTRec~\cite{ho2023btrec} and CR-SoRec~\cite{prakash2023cr} append a binary question towards the user preference after the textual description of user profile, user behaviors, and target item, and therefore convert the item scoring task into a binary question answering problem. 
Then, they can intercept the estimated score $s\in\mathbb{R}^V$ or probability $p\in\mathbb{R}^V$ in Eq.~\ref{eq:llm process} and conduct a bidimensional softmax over the corresponding logits of the binary key answer words (\ie, the token used to denote label, for example, Yes/No) for pointwise scoring:
\begin{equation}
    \hat{y}=\frac{\exp(p_{Yes})}{\exp(p_{Yes}+\exp(p_{No})}\in(0,1),
\end{equation}
where $p_{Yes}$ and $p_{No}$ denote the logits for ``Yes'' and ``No'' tokens, respectively.
Other typical examples that extract the softmax probabilities of corresponding label tokens for item scoring include TabLLM~\cite{hegselmann2023tabllm},
Prompt4NR~\cite{zhang2023prompt}, and
GLRec~\cite{wu2023exploring}. 
Moreover, another line of research intends to concatenate the item description (\eg, title) to the user behavior history with different templates, and estimates the score by calculating the overall perplexity~\cite{mao2023unitrec,qu2023thoroughly}, log-likelihood~\cite{sileo2022zero,sanner2023large}, or joint probability~\cite{zhang2021language} of the prompting text as the final predicted score $\hat{y}$ for user preference.
Besides, \citet{zhiyuli2023bookgpt} instruct LLM to predict the user rating in a textual manner, and restrict the output format as a value with two decimal places through manually designed prompts. 

\subsubsection{Item Generation Task}

In item generation tasks, the large language model serves as a generative function $F(u)$ to directly produce the final ranked list of items, requiring only one forward of function $F(u)$. 
Generally speaking, the item generation task highly relies on the intrinsic reasoning ability of LLM to infer the user preference and generate the ranked item list, the process of which can be formulated as:
\begin{equation}
    [i_k]_{k=1}^N = F(u), \; s.t.\; i_k \in \mathcal{I}.
\end{equation}
According to whether a set of candidate items is provided for LLM to accomplish the item generation task, we can categorize the related solutions into two classes: (1) open-set item generation, and (2) closed-set item generation.

In \textbf{open-set item generation} tasks~\cite{bao2023bi,salinas2023unequal,di2023evaluating,jiang2023reformulating,lin2023multi,mei2023lightlm,zheng2023adapting,qiu2023controlrec,liao2023llara,harte2023leveraging,li2023distillation,zhai2023knowledge,zhou2023exploring,li2023exploring,hua2023index,geng2023vip5,li2023gpt4rec,yin2023heterogeneous,jeong2023factual,hua2023up5}, LLM is required to directly generate the ranked item list that the user might prefer according to the user profile and behavior history \textit{without a given candidate item set}. 
Since the candidate items are not provided in the input prompt, the large language model is actually not aware of the universal item pool $\mathcal{I}$, thus bringing the \textit{generative hallucination problem}~\cite{mei2023lightlm}, where the generated items might fail to match the exact items in the item pool $\mathcal{I}$. 
Therefore, apart from the design of input prompt templates~\cite{hou2023large,li2023preliminary} and finetuning algorithms~\cite{li2023distillation}, the post-processing operations for item grounding and matching after the item generation are also required to overcome the generative hallucination problem~\cite{mei2023lightlm}.
We formulate the process as follows:
\begin{equation}
\begin{aligned}
    \left[\hat{i}_k\right]_{k=1}^N &\leftarrow \operatorname{LLM}(x_u),\\
    [i_k]_{k=1}^N &\leftarrow \operatorname{Match}\left(\left[\hat{i}_k\right]_{k=1}^N,\mathcal{I}\right),
\end{aligned}
\end{equation}
where $\operatorname{Match}(\cdot,\cdot)$ is the matching function, $\hat{i}_k$ is the LLM-generated items, and $i_k$ is the actual item matched from $\mathcal{I}$ according to $\hat{i}_k$.
LANCER~\cite{jiang2023reformulating} employs knowledge-enhanced prefix tuning for generation ground and further applies cosine similarity to match the encoded representation of generated item text with the universal item pool $\mathcal{I}$. 
\citet{di2023evaluating} leverage ChatGPT for user interest modeling and next item title generation with Damerau-Levenshtein distance~\cite{miller2009levenshtein} for item matching. 

Apart from generating the items in textual manners, another line of research focuses on aligning the language space with the ID-based recommendation space, and therefore enables LLM to generate the item IDs directly. 
For instance, \citet{hua2023index} explore better ways for item indexing (\eg, sequential indexing, collaborative indexing) in order to enhance the performance of such index generation tasks.
LightLM~\cite{mei2023lightlm} designs a lightweight LLM with carefully designed user \& item indexing, and applies constrained beam search for open-set item ID generation.
Besides, LLaRA~\cite{liao2023llara} represents items in LLM’s input prompts using a novel hybrid approach that integrates ID-based item embeddings from traditional recommenders with textual item features. 
Other typical works for open-set item generation include: 
GenRec~\cite{ji2023text}, TransRec~\cite{lin2023multi}, LC-Rec~\cite{zheng2023adapting}, ControlRec~\cite{qiu2023controlrec}, and POD~\cite{li2023distillation}.

In \textbf{closed-set item generation} tasks~\cite{ghosh2023jobrecogpt,wang2023multiple,yue2023llamarec,xu2023llms,yao2023knowledge,liu2023recprompt,wang2023drdt,singh2023scaling,sun2023large,chen2023palr,wang2023zero,hou2023large,zhang2023chatgpt,geng2023vip5,ma2023large,luo2024integrating,hua2023up5}, LLM is required to rank or select from \textit{a given candidate item set}. 
That is, we will first employ a lightweight retrieval model to pre-filter the universal item set $\mathcal{I}$ into a limited number of candidate items denoted as $\mathcal{C}=\{i_j\}_{j=1}^{J}, \,J \ll |\mathcal{I}|$. 
The number of candidate items is usually set up to 20 due to the context window limitation of LLM. 
The content of candidate items is then presented in the input prompt for LLM to generate the ranked item list, which can be formulated as:
\begin{equation}
\begin{aligned}
    \mathcal{C} &\leftarrow \operatorname{Pre-filter}(u,\mathcal{I}),\\
    [i_k]_{k=1}^N &\leftarrow \operatorname{LLM}(u,\mathcal{C}),\;N\le |\mathcal{C}|,
\end{aligned}
\end{equation}
For example, LlamaRec~\cite{yue2023llamarec} adopts LRURec~\cite{yue2023linear} as the retriever, and finetunes LLaMA2 for listwise ranking over the pre-filtered items. 
DRDT~\cite{wang2023drdt} ranks the given candidates with iterative multi-round reflection to to gradually refine the ranked list. 
LiT5~\cite{singh2023scaling} proposes to distill the zero-shot ranking ability from a proficient LLM (\eg, RankGPT4~\cite{sun2023chatgpt}) into a relatively smaller one (\eg, T5-XL~\cite{raffel2020exploring}).
AgentCF~\cite{zhang2023agentcf} incorporates LLM as the recommender by simulating user-item interactions in recommender systems through agent-based collaborative filtering.
Other typical examples include: JobRecoGPT~\cite{ghosh2023jobrecogpt}, InstructMK~\cite{wang2023multiple}, RecPrompt~\cite{liu2023recprompt}, PO4ISR~\cite{sun2023large}, etc.

In comparison of these two tasks, open-set generation tasks generally suffer from the generative hallucination problem, where the generated items might fail to match the exact items in the universal item pool. 
Therefore, the post-generation matching function is heavily required, which increases the inference overhead and might even hurt the final recommendation performance, especially for scenarios with item texts that largely differ from the language distribution of LLM.
On the contrary, closed-set generation tasks use a lightweight retrieval model as the pre-filter to provide a clear set of candidate items, and therefore the large language model is able to mitigate the hallucination problem. 
However, the introduction of candidate items in the input prompt of LLM can cause other problems. 
Firstly, LLM cannot handle a large number of candidates (usually less than 20) due to the context window limitation, and the final recommendation performance can somehow be limited by the retrieval model (\ie, pre-filter). 
Moreover, \citet{ma2023large} and \citet{hou2023large} reveal that shuffling the order of candidate items in the prompt can affect the ranking output, leading to unstable recommendation results. 
The aforementioned issues of closed-set generation tasks intrinsically stem from the existence of candidate item set in the input prompt, which can be well solved in open-set generation tasks.
In summary, we can observe that the open-set and closed-set generation tasks have complementary strengths and weaknesses compared with each other. 
Hence, the choice between them in practical applications actually depends on specific situations and problems we meet in real-world scenarios. 


\subsubsection{Hybrid Task}

In hybrid tasks, the large language model serves in a multi-task manner, where both the item scoring and generation tasks could be handled by a single LLM through a unified language interface.
The basis for supporting this hybrid functionality is that large language models are inherent multi-task learners~\cite{brown2020language,ouyang2022training}.
P5~\cite{geng2022recommendation}, M6-Rec~\cite{cui2022m6} and InstructRec~\cite{zhang2023recommendation} tune the encoder-decoder models for better alignment towards a series of recommendation tasks including both item scoring and generation tasks via different prompting templates. 
RecRanker~\cite{luo2023recranker} combines the pointwise scoring, pairwise comparison and listwise ranking tasks to explore the potential of LLM for top-N recommendation.
BDLM~\cite{zhang2023bridging} bridges the information gap between the domain-specific models and the general large language models for hybrid recommendation tasks via an information sharing module with memory storage mechanism. 
UniMP~\cite{wei2024towards} builds a unified large foundation for multiple tasks in personalized systems, \eg, preference prediction, personalized search.
Other works~\cite{liu2023chatgpt,sun2023chatgpt,dai2023uncovering} manually design task-specific prompts to call a unified central LLM (\eg, ChatGPT API) to perform multiple tasks, including but not restricted to pointwise rating prediction, pairwise item comparison, and listwise ranking list generation. 
There also exist benchmarks (\eg, LLMRec~\cite{liu2023llmrec}, OpenP5~\cite{xu2023openp5}) that test the LLM-based recommenders on various recommendation tasks like rating prediction, sequential recommendation, and direct recommendation.

\subsection{LLM for User Interaction}
\label{sec:user interaction}

In many of practical applications, recommending is a one-turn interaction process, where the system monitors the user behaviors (\eg, click and purchase) over time and then presents a tailored set of relevant items in certain pre-defined situations. 
Such a one-turn interaction lacks effective and versatile ways to acquire user interests and detect the user’s current situation or needs in complex scenarios. 
To this end, the advent of large language models presents a promising alternative, by offering a more active and adaptive form of user interaction~\citet{wang2023generative,deng2023unimind,wang2022unicrs}. 
Instead of relying solely on the past user behaviors passively, LLM could engage in real-time interactions with the users to gather more nuanced natural language feedback about their preferences. 

In general, the user interaction based on LLM in recommendation is commonly formed as a multi-turn dialogue, which is covered in conversational recommender systems~\cite{zhou2020tgredial,deng2023unimind,wang2022recindial,wang2022unicrs}. 
During such a dialogue, LLM provides an unprecedented richness in understanding users' interests and requirements by integrating context in conversation and applying the extensive open-world knowledge. LLM can support a recommender to make highly relevant and tailored recommendations through eliciting the current preferences of user, providing explanations for the item suggestions, or processing feedback by users on the made suggestions~\cite{jannach2021survey_crs}. 
In other words, the introduction of large language models makes recommender systems more feasible and user-friendly in terms of user interaction.
Specifically, from the perspective of interactive content~\cite{li2018towards,zhou2020improving}, the modes of LLM-based user interaction can be categorized into (1) task-oriented user interaction, and (2) open-ended user interaction .

\subsubsection{Task-oriented User Interaction}
\label{sec:task-oriented user interaction}

The task-oriented user interaction~\cite{zhou2020tgredial,ren2023kecr,yang2022mese,zhang2023vricr,wang2022follow,deng2023unimind} supposes that the user has a clear intent and the recommender system needs to support the user's decision making process or assist the user in finding relevant items. 
To be specific, LLM is integrated as a component of the recommender system, specially aiming at analyzing user intentions.
As a typical work, TG-ReDial~\cite{zhou2020tgredial} proposes to incorporate topic threads to enforce natural semantic transitions towards the recommendation and develops a topic-guided conversational recommendation method. It deploys three BERT~\cite{devlin2018bert} modules to encode user profiles, dialogue history, and track conversation topics, respectively. Then, the encoded features are fed into a pre-set recommendation module to recommend items, followed by a GPT2~\cite{radford2019language} module to aggregate the encoded knowledge for response generation. 
After each turn, the results are gathered and will be used to support the next round of dialogue interaction, such as understanding changes in user interest and analyzing user feedback, etc. 
The subsequent works roughly follow a similar process for task-oriented user interaction. 
While earlier works attempt to manage the dialogue understanding and response generation with relatively small language models (\eg, BERT and GPT2), recent works start to incorporate billion-level large language models for better conversational recommendation and improving the satisfaction of user interaction. 
MuseChat~\cite{dong2023musechat} builds a multi-modal LLM based on Vicuna-7B~\cite{vicuna2023} to provide reasonable explanation for the music recommendation during the user dialogue.
\citet{liu2023conversational} leverage the complementary collaboration between conversational RS and LLM for e-commercial pre-sales dialogue understanding and generation. 
\citet{he2023large} construct a conversational recommendation dataset with more diverse textual contexts, and find that LLM is able to outperform finetuned traditional conversational recommenders in
zero-shot settings.
Other typical works for task-oriented user interaction include: 
MESE~\cite{yang2022mese}, KECR~\cite{ren2023kecr}, 
UniMIND~\cite{deng2023unimind}, VRICR~\cite{zhang2023vricr}, TCP~\cite{wang2022follow}.


\subsubsection{Open-ended User Interaction}

The task-oriented user interaction draws a strong assumption that the user engages in the recommender system with specific goals to seek certain items. 
Differently, the open-ended user interaction~\cite{wang2022barcor,wang2022recindial,wang2022unicrs,ram2023multitask,leszczynski2023talktw,wang2023rethinking} assumes that the user's intent is vague, and the system needs to gradually acquire user interests or guide the user through interactions (including topic dialogue, chitchat, QA, etc.) to achieve the goal of recommendation eventually. 
Consequently, the role of LLM for open-ended user interaction is no longer limited to a simple component for dialogue encoding and response generation as discussed in Section~\ref{sec:task-oriented user interaction}. 
Instead, LLM plays a key role in driving the interaction process by leading and acquiring the user interests for final recommendation.
Specifically, BARCOR~\cite{wang2022barcor} proposes a unified framework based on BART~\cite{lewis2020bart} to first conduct user preference elicitation, and then perform response generation with recommended items, which aims to maximize the mutual information between conversation interaction and item recommendation. 
T5-CR~\cite{ram2023multitask} focuses on user interaction modeling and formulates conversation recommendation as a language generation problem. It adopts T5~\cite{raffel2020t5} to achieve dialogue context understanding, user preference elicitation, item recommendation and response generation in an end-to-end manner. Specifically, it adopts a special token symbol as the trigger to generate recommended item during the response generation.
\citet{wang2023rethinking} investigate the ability of ChatGPT to converse with user for item recommendation and explanation generation through manually designed prompts without any demonstration (\ie, zero-shot prompting). Then, they utilize LLM as an auxiliary user interaction component for dialogue understanding and user preference elicitation. 
Other related research works include:
UniCRS~\cite{wang2022unicrs}, RecInDial~\cite{wang2022recindial}, and TtW~\cite{leszczynski2023talktw}.

\subsection{LLM for Pipeline Controller}

As the model size scales up, LLM tends to exhibit emergent behaviors that may not be observed in previous smaller language models, \eg, few-shot in-context learning, instruction following, step-by-step reasoning, and tool usage~\cite{wei2022emergent,zhao2023survey}.
With such emergent abilities, LLM is no longer just a part of the recommender system mentioned above, but could actively participate in the pipeline control over the system, possibly leading to a more interactive and explainable recommendation process~\cite{jin2023lending}. 
Chat-REC~\cite{gao2023chat} leverages ChatGPT to bridge the conversational interface and traditional recommender systems, where it is required to infer user preferences, decide whether or not to call the backend recommendation API, and further modify (\eg, filter and rerank) the returned item candidates before presenting them to the user. These operations enable LLM to step beyond the role for user interaction in Section~\ref{sec:user interaction}, and cast controls for the multi-stage recommendation pipeline with certain API calls and tool usage for conversational recommender systems.
RecLLM~\cite{friedman2023leveraging} further extends the permission of LLM, and proposes a roadmap for building an integrated conversational recommender system, where LLM is able to manage the dialogue, understand user preference, arrange the ranking stage, and even provide a controllable LLM-based user simulator to generate synthetic conversations.
RAH~\cite{shu2023rah} designs the Learn-Act-Critic loop and a reflection mechanism for LLM-based agents to improve the alignment with user preferences during the interaction period.
InteRecAgent~\cite{huang2023recommender} serves as the interactive agent for conversational recommendation with the users, and is accessible to a range of plug-in tools including but not limited to intention detection, information query, item retrieval, and item ranking. 
Besides, instead of allowing LLM to take over control of the entire recommendation pipeline, RecMind~\cite{wang2023recmind} makes finer-grained control over the recommendation process with task deconstruction. It proposes to address the user queries under self-inspiring prompting strategy and multi-step reasoning with tool usage (\eg, expert models, SQL tool, search engine).

\subsection{Discussion}
\label{sec:where discussion}

We could observe that the development path about where to adapt LLM to RS is fundamentally aligned with the progress of large language models.
Back in the year 2021 and early days of 2022, the parameter sizes of pretrained language models are still relatively small (\eg, 110M for BERT-base, 1.5B for GPT2-XL). 
Therefore, earlier works usually tend to either incorporate these small-scale language models as simple textual feature encoders, or as scoring/ranking functions finetuned to fit the data distribution of recommender systems. 
In this way, the recommendation process is simply formulated as a one-shot straightforward predictive task, and can be better solved with the help of language models.

As the model size gradually increases, researchers discover that large language models have gained emergent abilities (\eg, instruction following and reasoning), as well as a vast amount of open-world knowledge with powerful text generation capacities.
Equipped with these amazing features brought by large-scale parameters, LLM starts to not only deepen its usage in the feature encoder and scoring/ranking function stage, but also further extend their roles into other stages of the recommendation pipeline.
For instance, in the feature engineering stage, we could instruct LLM to generate reliable auxiliary features and synthetic data samples~\cite{liu2023first} to assist the model training and evaluation. In this way, the open-world knowledge from LLM is injected into the closed-domain recommendation models. 
Furthermore, large language models also revolutionize the user interaction with a more human-friendly natural language interface and free-form dialogue for various information systems.
Not to mention, participating in the pipeline control further requires sufficient logical reasoning and tool utilization capabilities, which are possessed by large language models.

In summary, we believe that, as the abilities of large language models are further explored, they will form gradually deeper couplings and bindings with multiple stages of the recommendation pipeline. 
Even further, we might need to customize large language models specifically tailored to satisfy the unique requirements of recommender systems~\cite{lin2023sparks}.

%% file: sections/how-to-adapt-LLMs.tex
    \section{How to Adapt Large Language Models}
\label{sec: how}

\begin{figure}[t]
    \centering
    \includegraphics[width=0.9\textwidth]{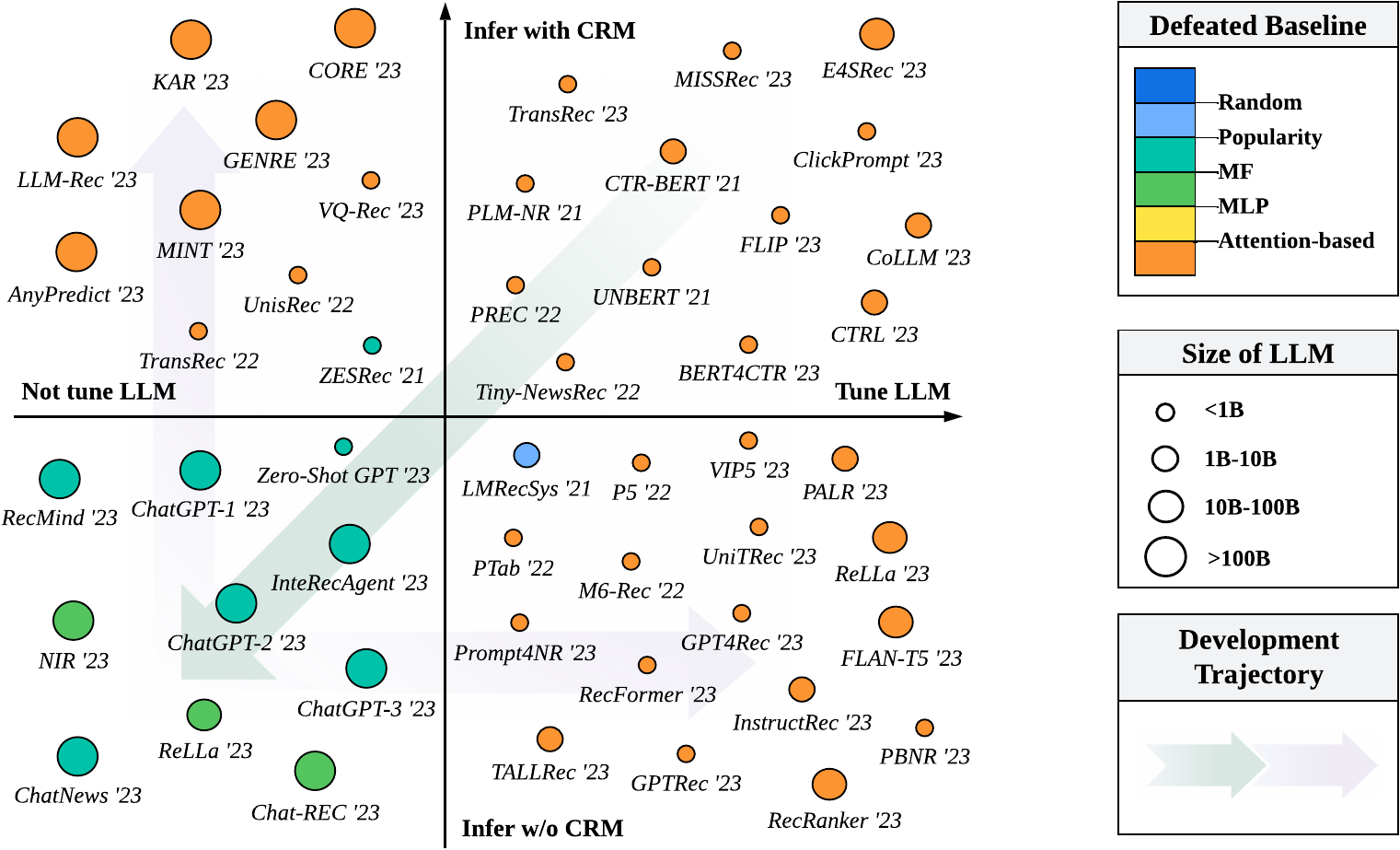}
    \caption{Four-quadrant classification about how to adapt LLM to RS. 
    Each circle in the quadrants denotes one research work with the corresponding model name attached below the circle.
    The size of each circle means the largest size of LLM leveraged in the research work.
    The color of each circle indicates the best compared baseline that the proposed model defeats as reported in the corresponding paper. 
    For example, the \textbf{\textcolor[rgb]{0.3294,0.7686,0.3686}{green}} circle of \textbf{Chat-REC} in quadrant 3 denotes that it utilizes a large language model with size larger than 100B (\ie, ChatGPT) and defeats the MF baseline.
    Besides, we summarize the general development path with light-colored arrows.
    Abbreviations: MF is short for matrix factorization; MLP is short for multi-layer perceptron.
    }
    \label{fig:how LLM RS}
\end{figure}


To answer the \textbf{``HOW''} question about adapting LLM to RS, we carry out two orthogonal taxonomy criteria to distinguish the adaptation of LLM to RS, resulting in a four-quadrant classification shown in Figure~\ref{fig:how LLM RS}:
\begin{itemize}[leftmargin=10pt]
    \item \textbf{Tune/Not Tune LLM} denotes whether we will tune LLM based on the in-domain recommendation data during the training phase. The definition of tuning LLM includes both full finetuning and other parameter-efficient finetuning methods (\eg, LoRA~\cite{hu2021lora}, prompt tuning~\cite{lester2021power}).
    \item \textbf{Infer with/without CRM} denotes whether we will involve conventional recommendation models (CRM) during the inference phase. Note that there are works that only use CRM to serve as independent pre-filter functions to generate the candidate item set for LLM~\cite{ghosh2023jobrecogpt,wang2023multiple,yue2023llamarec}. We categorize them as ``infer without CRM'', since the CRM is independent of LLM, and could be decoupled from the final recommendation task.
\end{itemize}

In Figure~\ref{fig:how LLM RS}, we use different marker sizes to indicate the size of the large language model the research works adapt, and use different colors to indicate the best baseline they have defeated in terms of item recommendation.
Thus, a few works are not presented in Figure~\ref{fig:how LLM RS} since they do not provide traditional recommendation evaluation, \eg, RecLLM~\cite{friedman2023leveraging} only investigates the system architecture design to involve LLM for RS pipeline control without experimental evaluation. 
Moreover, it is noteworthy that some research works might propose techniques that are applied across different quadrants. For instance, ReLLa~\cite{lin2023rella} designs semantic user behavior retrieval to help LLM better comprehend and model the lifelong user behavior sequence in both zero-shot prediction (\ie, quadrant 3) and few-shot finetuning (\ie, quadrant 4) settings.

Given the four-quadrant taxonomy, we demonstrate that the overall development path in terms of ``\textbf{HOW}'' research question generally follows the light-colored arrows in Figure~\ref{fig:how LLM RS}.
Accordingly, we will introduce the latest research works in the order of quadrant 1, 3, 2, 4, followed by in-detail discussions for each quadrant subsection.

\subsection{Tune LLM \,\&\, Infer with CRM (Quadrant 1)}

Quadrant 1 refers to research works that not only finetune the large language models with in-domain recommendation data during the training phase, but also introduce conventional recommendation models to provide better collaborative knowledge. 
Based on their development over time, the works in quadrant 1 can be mainly classified into two stages.

Back to years 2021 and 2022, earlier works in quadrant 1 mainly focus on applying relatively smaller pretrained language models (\eg, BERT~\cite{devlin2018bert} and GPT2~\cite{radford2019language}) to the downstream domains with abundant textual features, \eg, news recommendation~\cite{zhang2021unbert,wu2021empowering,liu2022boosting,yu2022tiny}, web search~\cite{zou2021pre,liu2021pre} and e-commercial advertisement~\cite{muhamed2021ctr,wang2022learning}. 
As discussed in Section~\ref{sec:where discussion}, the primary roles of these small-scale language models are only limited to feature encoders for semantic representation enhancement. Consequently, a conventional recommendation model (CRM) is required to make the final recommendation, with generated textual representations as auxiliary inputs.
Additionally, the small model size makes it affordable to fully finetune the language model during the training phase. 
UNBERT~\cite{zhang2021unbert}, PLM-NR~\cite{wu2021empowering}, PREC~\cite{liu2022boosting} and Tiny-NewsRec~\cite{yu2022tiny} conduct full finetuning over small-scale language models (\eg, BERT~\cite{devlin2018bert}, RoBERTa~\cite{liu2019roberta}, UniLMv2~\cite{bao2020unilmv2}) to enhance the content understanding for better news recommendation with sequential CRMs.
\citet{zou2021pre} and \citet{liu2021pre} customize pretrained language models with further pretraining over the domain-specific corpus for web search scenarios.
Moreover, TransRec~\cite{fu2023exploring} proposes layerwise adapter tuning over BERT~\cite{devlin2018bert} to ensure both the training efficiency and multi-modality enhanced representations. 
Although these earlier works can defeat strong baseline models with attention mechanisms by tuning language models and involving CRM, they only leverage small-scale language models as feature encoders, and thus the key capacities (\eg, reasoning, instruction following) of large foundation models remain underexplored.

At around the beginning of the year 2023, the rise of LLM (\eg, ChatGPT) demonstrates impressive emergent abilities like reasoning and instruction following, pointing out promising directions for LLM-enhanced recommendation.
Therefore, researchers start to investigate the potential of incorporating billion-level large language models (\eg, LLaMA~\cite{touvron2023llama} and ChatGLM~\cite{du2022glm,zeng2022glm}) to the field of recommender systems. 
Compared to earlier works with small-scale language models that we have discussed above, there are two major differences to be clarified for these recent works that incorporate large language models:
\begin{itemize}[leftmargin=10pt]
    \item Due to the massive amount of model parameters possessed by LLM, we can hardly perform full finetuning on LLM as it can lead to an unaffordable cost in computational resources. 
    Instead, parameter-efficient finetuning (PEFT) methods are commonly adopted for training efficiency with usually less than 1\% parameters need to be updated, \eg, low-rank adaption (LoRA)~\cite{hu2021lora} and prompt tuning~\cite{lester2021power,li2021prefix}. 
    \item The role of LLM is no longer a simple tunable feature encoder for CRM. 
    To make better use of the reasoning ability and open-world knowledge exhibited by LLM, researchers tend to place LLM and CRM on an equal footing (\eg, both as the recommenders), mutually leveraging their respective strengths to collaborate and achieve improved recommendation performance. 
    Moreover, as discussed in Section~\ref{sec: where}, LLM can also be finetuned for the stages of feature engineering~\cite{li2023ecomgpt}, user interaction~\cite{leszczynski2023talktw} and pipeline control~\cite{friedman2023leveraging} as well.
\end{itemize}
CoLLM~\cite{zhang2023collm} and E4SRec~\cite{li2023e4srec} adopt LoRA to finetune Vicuna-7B~\cite{vicuna2023} and LLaMA2-13B~\cite{touvron2023llama} respectively, and build personalized prompts by injecting the user \& item embedding from a pretrained CRM via a linear mapping layer.
CTRL~\cite{li2023ctrl} conducts knowledge distillation between LLM and CRM for better alignment and interaction between the semantic and collaborative knowledge, where the size of involved LLM scales up to 6 billion (ChatGLM-6B~\cite{zeng2022glm}) with last-layer finetuning strategy. 
LLaMA-E\cite{shi2023llama} and EcomGPT~\cite{li2023ecomgpt} finetune the base large language models (\ie, LLaMA-30B~\cite{touvron2023llama} and BLOOMZ-7.1B~\cite{muennighoff2022crosslingual}) to assist the conventional recommendation models with augmented generative features, \eg, item attributes and topics of user reviews. 

As shown in Figure~\ref{fig:how LLM RS}, since CRM is involved and LLM is tunable, the research works in quadrant 1 could better align to the data distribution of recommender systems and thus all achieve satisfying performance, even when the size of involved LLM is relatively small. 
Moreover, we can observe the clear trend that researchers intend to consider larger language models from the million level up to the billion level, thus benefiting from their vast amount of open-world semantic knowledge, as well as the instruction following and reasoning abilities.
Nevertheless, when it comes to low-resource scenarios, the small-scale language model (\eg, BERT) is still an economic choice to balance between LLM-based enhancement and computational efficiency.


\subsection{Not Tune LLM \,\&\, Infer w/o CRM (Quadrant 3)}
\label{sec:quadrant 3} 

Quadrant 3 refers to research works that exclude the conventional recommendation model and solely adopt a frozen large language model as the recommender. 
This line of research generally emerges since the advent of large foundation models, especially ChatGPT, where researchers aim to analyze the zero-shot or few-shot performance of LLM in recommendation domains with LLM fixed and CRM excluded. 
It should be noted that, in the context of quadrant 3, the ``few-shot'' setting specifically refers to the in-context learning (ICL) approach for LLM, rather than tuning LLM based on a few training samples.

Earlier works~\cite{wang2023zero,liu2023chatgpt,sun2023chatgpt,dai2023uncovering} investigate the zero-shot and few-shot recommendation settings based on the ChatGPT API, with delicate prompt engineering to instruct the LLM to model the user interest and perform tasks like rating prediction, pairwise comparison, and listwise ranking.
However, the performance of these approaches is not satisfactory.
Based on these previous works, several attempts~\cite{lin2023rella,gao2023chat,zhang2023agentcf,wang2023recmind} are made to improve the zero-shot or few-shot recommendation performance of LLM.
\citet{lin2023rella} identify the lifelong sequential behavior incomprehension problem of LLM, \ie, LLM fails to extract the useful
information from a textual context of long user behavior sequence for recommendation tasks, even if the length of context is far from reaching the context limitation of LLM. 
To mitigate such an issue, ReLLa~\cite{lin2023rella} proposes to perform semantic user behavior retrieval to replace the simply truncated top-$K$ recent behaviors with the top-$K$ semantically relevant behaviors towards the target item. 
In this way, the quality of data samples is improved, thus making it easier for LLM to comprehend the user sequence and achieve better zero-shot recommendation performance. 
RecMind~\cite{wang2023recmind} designs the self-inspiring prompt strategy and enables LLM to explicitly access the external knowledge with extra tools, such as SQL for recommendation database and search engine for web information.
Chat-REC~\cite{gao2023chat} instructs ChatGPT to not only serve as the score/ranking function, but also take control over the recommendation pipeline, \eg, deciding when to call an independent pre-ranking model API. 

As illustrated in Figure~\ref{fig:how LLM RS}, although a larger model size might bring performance improvement, the zero-shot or few-shot learning of LLM in quadrant 3 is much inferior compared with the light-weight CRM tuned on the training data. 
Even when equipped with advanced techniques such as user behavior retrieval and tool usage, the performance of a frozen LLM without CRM is still suboptimal and far from the SOTA performance. 
The knowledge contained in LLM is global and factual, but recommendation is a personalized task that requires preference-oriented knowledge.
This indicates the importance of in-domain collaborative knowledge from the training data of recommender systems, and that solely relying on a fixed large language model is currently unsuitable to well tackle the recommendation tasks. 
Consequently, there are two major approaches to further inject the in-domain collaborative knowledge for LLM to improve the recommendation performance: (1) involving CRM for inference, and (2) tuning LLM based on the training data, which refer to works of quadrant 2 and quadrant 4 in Figure~\ref{fig:how LLM RS}, respectively. 


\subsection{Not Tune LLM \,\&\, Infer with CRM (Quadrant 2)}

Research works in quadrant 2 utilize different key capabilities (\eg, rich semantic information, reasoning ability) of LLM without finetuning to help CRM better accomplish the recommendation tasks. 
Similar to works in quadrant 1, the utilization of a frozen LLM in quadrant 2 generally demonstrates a development path in terms of the size of LLM evolving over time, \ie, from small-scale language models to large language models.

Early works~\cite{ding2021zero,hou2022towards,hou2023learning} propose to extract transferable text embeddings from a fixed BERT~\cite{devlin2018bert} model with rich semantic information. 
The text embeddings are then fed into several projection layers to better produce the cross-domain representations as the input of trainable conventional recommendation models.
The projection layers are designed as a single-layer neural network for ZESRec~\cite{ding2021zero}, a mixture-of-expert (MoE) network for UniSRec~\cite{hou2022towards}, and a vector quantization based embedding lookup table for VQ-Rec~\cite{hou2023learning}. 
We can observe from Figure~\ref{fig:how LLM RS} that the direct usage of a single-layer neural network as an adapter does not yield satisfactory results. However, with a carefully designed adapter module, the semantic representations from the fixed BERT parameters can be better aligned with the subsequent recommendation module, leading to impressive recommendation performances. 

As discussed in Section~\ref{sec:where discussion}, with the model size scaling up, the emergent abilities and abundant open-world knowledge enable large foundation models to extend their roles to other stages of the recommendation pipeline, such as feature engineering stage~\cite{xi2023towards,liu2023first,wang2023anypredict,li2023taggpt} and user interaction~\cite{hou2022towards,ren2023kecr,wang2023rethinking,he2023large}. 
AnyPredict~\cite{wang2023anypredict} leverages ChatGPT APIs to consolidate tabular samples to overcome the barrier across tables with varying schema, resulting in unified expanded training data for the follow-up conventional predictive models.
ONCE~\cite{liu2023first} utilizes ChatGPT to perform news piece generation, user profiling, and news summarization, and thus augments the news recommendation model with LLM-generated features.
KAR~\cite{xi2023towards} and RLMRec~\cite{ren2023representation} leverage LLM to enhance the user behavior modeling with specially designed input templates as well as chained prompting strategies, aiming to provide user-level feature augmentation for CRM. 
\citet{wang2023rethinking} and \citet{he2023large} investigate the integration of LLM (\eg, ChatGPT and GPT4) to handle the open-ended free-form chatting during the user interaction of conversational recommendation.

In these works, although LLM is frozen, the involvement of CRM for the inference phase generally guarantees better recommendation performance, compared with works from quadrant 3 (\ie, Not Tune LLM; Infer w/o CRM) in terms of the best baseline they defeat. 
When compared with quadrant 1 (\ie, Tune LLM; Infer with CRM), since the large language model is fixed, the role of LLM in quadrant 2 is mostly auxiliary for CRM at different stages of the recommendation pipeline, including but not limited to feature engineering and feature encoder. 

\subsection{Tune LLM \,\&\, Infer w/o CRM (Quadrant 4)}
\label{sec:how quadrant 4}

Research works in quadrant 4 aim to finetune the large language models to serve as the scoring/ranking function based on the training data from recommender systems, excluding the involvement of CRM. 
Since CRM is excluded, we have to apply prompt templates to obtain textual input-output pairs, and therefore convert the recommendation tasks (\eg, click-through rate estimation and next item prediction) into either a text classification task~\cite{bao2023tallrec,lin2023rella} or a sequence-to-sequence task~\cite{geng2022recommendation,wang2023multiple,singh2023scaling,wei2024towards}.

As an early attempt, LMRecSys~\cite{zhang2021language} tunes language models to estimate the score of each candidate item via joint inference over multiple masked tokens, resulting in unsatisfying performance.
The reason might be that its scoring manners are somehow problematic, where the authors simply pad or truncate the length of all the titles for items to 10 tokens.
Prompt4NR~\cite{zhang2023prompt} finetunes BERT by predicting the key answer words (\eg, Yes/No, Good/Bad) based on the prompting templates. 
PTab~\cite{liu2022ptab} transforms tabular data into text and finetunes a BERT model based on the masked language modeling task followed by classification tasks.
UniTRec~\cite{mao2023unitrec} finetunes a BART~\cite{lewis2020bart} model with a joint contrastive loss to optimize the discriminative score and a perplexity-based score. 
RecFormer~\cite{li2023text} adopts two-stage finetuning based on masked language modeling loss and item-item contrastive loss with LongFormer~\cite{beltagy2020longformer} as the backbone model.  
P5~\cite{geng2022recommendation}, FLAN-T5~\cite{kang2023llms}, and InstructRec~\cite{zhang2023recommendation} adopt T5~\cite{raffel2020exploring} as the backbone, and train the model in a sequence-to-sequence manner.
GPT4Rec~\cite{li2023gpt4rec} tunes GPT~\cite{radford2019language} models as a generative function for next item prediction via causal language modeling.

The works mentioned above all adopt full finetuning over relatively small-scale language models (\eg, 110M for BERT-base, 149M for LongFormer), which could be considerably expensive and unscalable as the size of the language model continuously increases up to tens of or even hundreds of billions. 
Although the large model parameter capacity enables proficient knowledge and capabilities, fully finetuning such a big model can lead to substantial resource consumption.
As a result, parameter-efficient finetuning methods (PEFT) are usually required to efficiently adapt billion-level LLM to RS. 
Among those PEFT methods, low-rank adaption (LoRA) serves as the most popular choice. 
For instance, ReLLa~\cite{lin2023rella}, GenRec~\cite{ji2023genrec}, BIGRec~\cite{bao2023bi}, RecSysLLM~\cite{chu2023leveraging} and LSAT~\cite{shi2023preliminary} adopt the LoRA~\cite{hu2021lora} technique to finetune a base large language model (usually LLaMA-7b~\cite{touvron2023llama} or Vicuna-7B~\cite{vicuna2023}) for item scoring or generation tasks. 
Apart from LoRA, M6-Rec~\cite{cui2022m6} designs option tuning as an improved version of prompt tuning to empower M6~\cite{lin2021m6} for varied downstream tasks like item retrieval and ranking.
VIP5~\cite{geng2023vip5} performs layerwise adapter tuning to unify various modalities (\eg, ID, text, and image) via a universal foundation model for recommendation tasks. 

Although the introduction of PEFT alleviates the training inefficiency issue, the computational overhead can still be excessive for real-world applications where the number of training samples might scale up to billions. 
In such a situation, even PEFT methods like LoRA are not efficient enough for LLM to go over the entire training dataset. 
To this end, recent works start to investigate the strong inductive learning capability~\cite{prince2006inductive} of LLM by downsampling the whole training set into a small-scale subset~\cite{lin2023rella,chen2023palr,kang2023llms,luo2023recranker}. 
As a representative, ReLLa~\cite{lin2023rella} uniformly samples less than 10\% of the training instances and surprisingly finds that LLM, which is finetuned only based on less than 10\% samples, is able to outperform the conventional recommendation baseline models that are trained on the entire training dataset. 
Such a phenomenon about the strong few-shot inductive learning capability of LLM in recommendation is also validated by other related works~\cite{chen2023palr,kang2023llms,luo2023recranker}.
As for different downsampling strategies, PALR~\cite{chen2023palr} randomly selects 20\% of the user to construct the training subset for efficient finetuning of LLaMA-7B~\cite{touvron2023llama}. 
RecRanker~\cite{luo2023recranker} designs an adaptive user sampling strategy, which consists of both importance-aware and clustering-based sampling followed the repetitive penalty.

As shown in Figure~\ref{fig:how LLM RS}, the performance of finetuning LLM based on recommendation data is promising with proper task formulation, even if the model size is still relatively small (\ie, less than 1 billion). 
Apart from the design of input prompt and model architecture to achieve superior recommendation performance, scalability and efficiency are also the major challenges in this line of research. 
That is, how to efficiently finetune a \textit{large-scale} language model on a \textit{large-scale} training dataset, where various PEFT methods and data downsampling strategies would be considered.

\subsection{Discussion}
\label{sec:how discussion}

We first conclude the necessity of collaborative knowledge injection when adapting LLM to RS, and then summarize the overall development path in terms of the ``\textbf{HOW}'' question, as well as possible future directions.
Next, we cast a discussion on the relationship between the recommendation performance and the size of the adapted LLM. 
Finally, we discuss an interesting property found about the hard sample reranking for large language models.

\subsubsection{Collaborative Knowledge is Needed}

From Figure~\ref{fig:how LLM RS}, we could observe a clear performance boundary between works from quadrant 3 and quadrant 1, 2, 4. 
The research works from quadrant 3 are inferior even though they adapt large-scale models (\ie, ChatGPT or GPT4), even when they are equipped with advanced techniques like user behavior retrieval and tool usage.
This indicates that the recommender system is a highly specialized area, which demands a lot of in-domain collaborative knowledge.
LLM cannot effectively learn such knowledge from its general pretraining corpus.
Therefore, we have to involve in-domain collaborative knowledge for better performance when adapting LLM to RS, and there are generally two ways to achieve the goal (corresponding to quadrant 1, 2, 4):
\begin{itemize}[leftmargin=10pt]
    \item \textbf{Tune LLM} during the training phase, which injects collaborative knowledge from a data-centric aspect.
    \item \textbf{Infer with CRM} during the inference phase, which injects collaborative knowledge from a model-centric aspect.
\end{itemize}

\begin{figure}[t]
    \centering
    \includegraphics[width=0.99\textwidth]{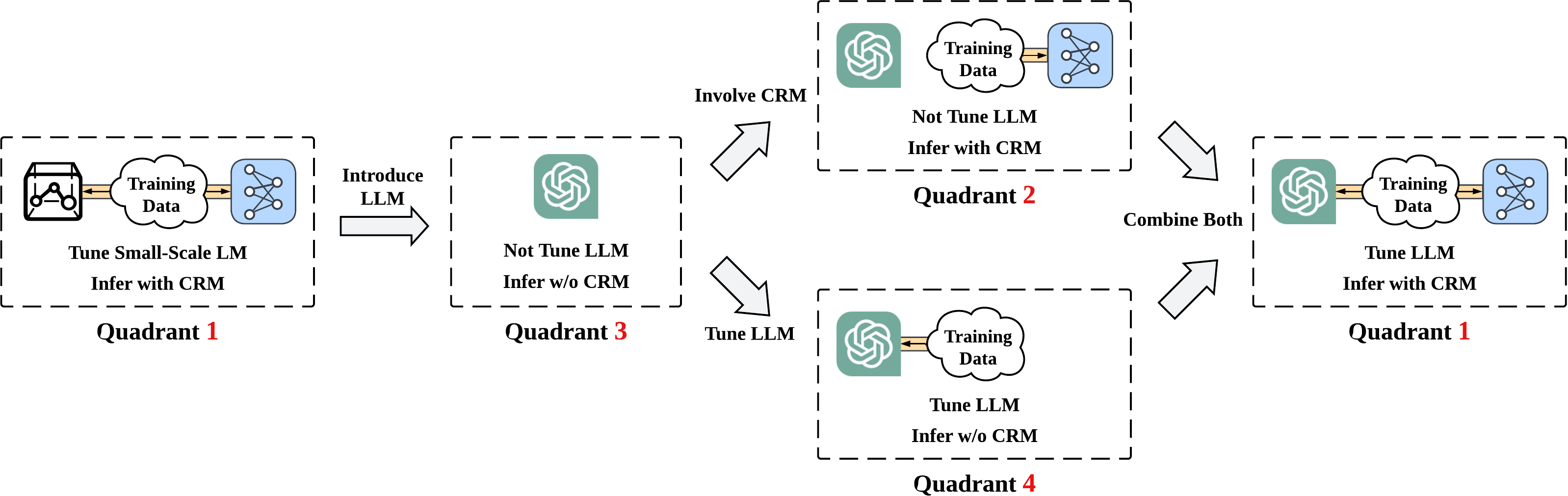}
    \caption{The illustration of the development trend for adapting LLM to RS in terms of the ``\textbf{HOW}'' research question based on the four-quadrant classification. 
    Earlier attempts generally perform joint optimization of small-scale language models and conventional recommendation models based on the training data (\ie, Quadrant 1). 
    Then, researchers try to introduce a frozen LLM for recommendation without the help of CRM (\ie, Quadrant 3), which results in inferior performance. 
    To this end, the golden principle, \ie, in-domain collaborative knowledge injection, is discovered, and a wide range of works start to explore the potential of LLM for RS by involving CRM (\ie,  Quadrant 2), tuning LLM (\ie, Quadrant 4), or combining both strategies (\ie, back to Quadrant 1).
    }
    \label{fig:how development}
\end{figure}

Both two approaches emphasize the importance of in-domain collaborative knowledge when adapting LLM to RS. 
Based on the insights above, as shown in Figure~\ref{fig:how development}, we draw a general development trend about the ``\textbf{HOW}'' research question on the basis of the four-quadrant classification. 
Starting from the early days of the year 2021, researchers usually intend to combine both small-scale LM and CRM to conduct joint optimization for recommendation (\ie, Quadrant 1). 
Then, at around the beginning of the year 2023, several works begin to leverage a frozen LLM for recommendation without the help of CRM (\ie, Quadrant 3), the inferior performance of which indicates the necessity of collaborative knowledge. 
To this end, two major solutions are proposed to conduct the in-domain collaborative knowledge injection via either involving CRM or tuning LLM, corresponding to Quadrants 2 and 4, respectively. 
Next, as we discover the golden principle for the adaptation of LLM to RS (\ie, in-domain collaborative knowledge injection), the development path further moves back to Quadrant 1, where we aim to jointly optimize LLM and CRM for superior recommendation performance. 
Finally, in terms of how to adapt LLM to RS, the possible future direction might lie in the ways to better incorporate the collaborative knowledge from recommender systems with the general-purpose semantic knowledge and emergent abilities exhibited by LLM. 
For example, empowering agent-based LLM with external tools for more thorough access to recommendation data, as well as real-time web information from search engines.

\subsubsection{Is Bigger Always Better?}

By injecting in-domain collaborative knowledge from either data-centric or model-centric aspects, research works from quadrants 1, 2, and 4 can achieve satisfying recommendation performance compared with attention-based baselines, except for a few cases.
Among these studies, although we could observe that the size of adapted LLM gradually increases according to the timeline, a fine-grained cross comparison among them (\ie, a unified benchmark) remains vacant.
Hence, it is difficult to directly conclude that a larger model size of LLM can definitely yield better results for recommender systems. 
This gives rise to an open question: \emph{Is bigger language models always better for recommender systems? Or is it good enough to use small-scale language models in combination with collaborative knowledge injection?}
Our opinions towards the question are in two folds:
\begin{itemize}[leftmargin=10pt]
    \item Compared with small-scale language models, large language models are still irreplaceable in certain specific tasks where reasoning abilities are required. For example, textual feature augmentation, human-like user interaction \& dialogue, and recommendation pipeline control. 
    In these scenarios, it is usually necessary to involve LLM instead of small-scale LM to ensure task accomplishment and recommendation performance.
    \item When playing the same role in RS (\eg, feature encoder), it is generally a commonsense that LLM can achieve better performance than small-scale LM. 
    However, small-scale LM would serve as a more economical choice to balance between performance enhancement and computational cost. 
    Or to say, whether the additional computational cost brought by LLM is worth the performance gain is still not well verified, especially when having small-scale LM as the light-weight substitute.
\end{itemize}

\subsubsection{LLM is Good at Reranking Hard Samples}

Although LLM generally suffers from inferior performance for zero/few-shot learning since little in-domain collaborative knowledge is involved, researchers~\cite{ma2023large,hou2023large} have found that  large language models such as ChatGPT are more likely to be a good reranker for \emph{hard samples}.
They introduce the filter-then-rerank paradigm which leverages a pre-ranking function from traditional recommender systems (\eg, matching or pre-ranking stage in industrial applications) to pre-filter those easy negative items, and thus generates a set of candidates with harder samples for LLM to rerank. In this way, the listwise reranking performance of LLM (especially ChatGPT-like APIs) could be promoted.
This finding is instructive for industrial applications, where we could require LLM to only handle hard samples and leave other samples for light-weight models to save computational costs.

%% file: sections/challenge.tex
\section{Challenges from Real-world Applications}
\label{sec: challenge}

In this section, we highlight the key challenges in adapting LLM to RS, which mainly arise from the unique characteristics of recommender systems and real-world applications.
Accordingly, we will also discuss the preliminary efforts done by existing works, as well as other possible solutions. 
The following challenges are proposed from three aspects: (1) \textbf{efficiency} (training efficiency, and inference latency), (2) \textbf{effectiveness} (in-domain long text modeling, and ID indexing \& modeling), and (3) \textbf{ethics} (fairness, and other potential risks from LLM). 

\subsection{Training Efficiency}

There are two key aspects to improve the performance of modern deep learning based recommender systems: (1) enlarge the volumes of training data (\eg, billion-level training samples), and (2) increase the update frequency for model (from day-level to hour-level, or even minute-level).
Both of these two factors highly require the training efficiency.
Although, as suggested in Section~\ref{sec:how discussion}, tuning LLM (possibly with CRM) is a promising approach to align LLM to RS for better performance, it actually brings prohibitive adaptation costs in terms of both time and computational resource consumption.
Therefore, how to ensure the training efficiency when adapting LLM to RS is one of the key challenges for real-world applications.

Existing works mainly propose to leverage parameter-efficient finetuning (PEFT) methods (\eg, low-rank adaptation~\cite{hu2021lora}, option tuning~\cite{cui2022m6}, layerwise adapter tuning~\cite{geng2023vip5}), which mainly solve the memory usage problem, but the time consumption is still high especially for large-scale scenarios with massive users and items. 
From the perspective of real-world applications, we suggest adopting the asynchronous update strategy, when we leverage LLM for feature engineering and feature encoder. 
To be specific, we can cut down the training data volume and relax the update frequency for LLM (\eg, week-level) while maintaining full training data and high update frequency for CRM.
The basis to support this approach is that researchers~\cite{chen2023palr,zhou2023lima,lin2023rella} point out that LLM has strong inductive learning capacities to produce generalized and reliable outputs via a handful of supervisions. 
In this way, LLM can provide aligned in-domain knowledge to CRM, while CRM acts as a frequently updated adapter for LLM.

\subsection{Inference Latency}

Online recommender systems are usually real-time services and extremely time-sensitive, where all the stages (\eg, matching, ranking, reranking) should be done within around tens of milliseconds.
The involvement of LLM during the inference phase gives rise to the inference latency problem.
The inference time of the LLM is fairly high, not to mention the additional time cost brought by prompt template generation.

\emph{Pre-computing and caching} the outputs or middle representations of LLM serves as a common strategy to ensure low-latency inference when we have to involve LLM during the inference phase.
When adapting the LLM as the scoring/ranking function, M6-Rec~\cite{cui2022m6} proposes the multi-segment late interaction strategy.
The textual features of user and item are split into finer-grained segments that are more static, \eg, by representing each clicked
item as an individual segment.
Then, we can pre-compute and cache the encoded representations of each segment using the first several transformer layers, while the rest of the layers are leveraged to perform late interaction among multiple segments when the recommendation request arrives.
Other works like UniSRec~\cite{hou2022towards} and VQ-Rec~\cite{hou2023learning} simply adopt language models as feature encoders. Hence it is straightforward to directly cache the dense embeddings produced by the language model. 
The pre-computing and caching strategy might be suitable for item-side information since they are generally static, but it can be suboptimal for user-side information since the user behaviors and interests are highly dynamic and quickly evolve over time. 
Hence, we have to find an appropriate caching frequency to balance between the performance and computational cost.

Moreover, we can also seek ways to \emph{reduce the size of model} for the inference efficiency, where methods have been well explored in other deep learning domains, \eg, distillation~\cite{jiao2019tinybert}, pruning~\cite{chen2020lottery}, and quantization~\cite{zafrir2019q8bert}.
For instance, CTRL~\cite{li2023ctrl} and FLIP~\cite{wang2023flip} propose to perform contrastive learning to distill the semantic knowledge from LLM to CRM. The CRM is then solely finetuned with improved parameter initialization for better recommendation performance, concurrently maintaining the low-latency inference.
These strategies generally involve a tradeoff between the model performance and inference latency.
Alternatively, we could involve LLM in the feature engineering stage and pre-store the outputs of LLM, which will bring a significantly smaller (but not entirely negligible) extra burden for inference.
Besides, we can also introduce LLM to scenarios with relatively loose inference latency constraints like conversational recommender systems.

\subsection{In-Domain Long Text Modeling}

When adapting LLM, we have to construct in-domain textual inputs via prompting templates and insert proper instructions and demonstrations at the front if needed.
However, the general guideline of real-world recommender systems requires longer user history, larger candidate set and more features to achieve better recommendation performance, possibly leading to long-text inputs for LLM.
Such long-text inputs from RS domains (\ie, in-domain long texts) could result in two key challenges as follows.

Firstly, an excessively long-text input would cause the memory inefficiency problem (the space complexity of classical transformers are $O(L^2)$ where $L$ is the number of tokens), and might even break the context window limitation, leading to partial information lost and inferior outputs from LLM.
Secondly, even if the length of input prompt does not exceed the context window, there may still exist issues for LLM to fully comprehend and reason on the recommendation data. 
\citet{lin2023rella} and \citet{hou2023large} reveal that LLM has difficulty in dealing with long texts especially when we extend the text with longer user history or larger candidate set, even though the total number of input tokens is far from reaching the context window limitation (\eg, 512 for BERT, 4096 for ChatGPT).
The reason might be that the distribution of in-domain long text is quite different from the pretraining corpora of LLM.


To this end, it is of great importance to investigate how to properly filter, select, and arrange the textual information as the input for LLM during prompting engineering, as well as how to instruct or tune the LLM to better align with the distribution of these in-domain long-text inputs.
Besides, in NLP domains, a range of works are proposed to address the context window limitation (\eg, sliding windows~\cite{wang2019multi}, memory mechanism~\cite{ding2020cogltx}), which could be considered in recommender systems.
Moreover, recent works propose to combine the latent representations from CRM to compress the personalized input prompt for LLM, thus alleviating the long-text problem.
For instance, CoLLM~\cite{zhang2023collm} and E4SRec~\cite{li2023e4srec} replace the textual description of each user behavior with one latent vector mapping from the embedding table of CRM via a linear projection layer, which greatly reduces the number of tokens for long user sequences.
Inspired by prefix tuning~\cite{li2021prefix,liu2021p}, ClickPrompt~\cite{lin2023clickprompt} transforms the sample-wise final representation of CRM into layer-wise prompts for LLM, making it easier to eliminate unnecessary features from the prompt templates.



\subsection{ID Indexing \& Modeling}

In recommender systems, there exists a kind of pure ID features that inherently contains no semantic information (\eg, user ID, item ID).
If we include these ID features in the prompting text, the tokenization is actually unmeaningful to language models (\eg, user ID AX1265 might be tokenized as [AX, 12, 65]).
Many works~\cite{cui2022m6,hou2023learning} tend to directly abandon these ID features (\eg, replacing item IDs with item titles or descriptions) for unified cross-domain recommendation via the natural language interface, since the IDs are usually not shared across different domains. 
However, some works~\cite{geng2022recommendation,yuan2023go} point out that ID features can greatly promote the recommendation performance, although sacrificing the cross-domain generalization ability.
Therefore, it is still an open question about whether we should retain the ID features or not, which divides the research regarding ID indexing \& modeling into two directions.

On the one hand, we could sacrifice the cross-domain generalization ability to obtain better in-domain recommendation performance by keeping the ID features.
P5~\cite{geng2022recommendation} and its variants~\cite{geng2023vip5,hua2023up5,hua2023index} preserve the ID features as textual inputs in the prompting templates.
P5 designs a whole-word embedding layer to assign the same whole-word embedding for tokens from the same ID feature. The whole-word embeddings will be added to the token embeddings in the same way as position embeddings in language models.
Based on P5, \citet{hua2023index} further explore various item ID indexing strategies (\eg, sequential indexing, collaborative indexing) to ensure the IDs of similar items consist of similar sub-tokens.
RecFormer~\cite{li2023text} and UniSRec~\cite{hou2022towards} omit the item IDs in prompting texts, but introduce additional ID embeddings at either bottom embedding layer or top projection layer. 
Other works~\cite{zhang2023collm,li2023e4srec,zheng2023adapting} seek to integrate ID embeddings from conventional recommendation models and therefore make the input prompt of LLM free of pure ID features.
In summary, researchers in this line should focus on how to associate LLM with ID features via carefully designed ID indexing \& modeling strategies. 

On the other hand, we could abandon the ID features to achieve unified cross-domain recommendation via natural language interface.
Maintaining a unified model to serve various domains is very promising, especially when we involve large language model~\cite{cui2022m6,hou2023learning,lin2023rella}.
In this direction, in order to achieve similar performance to those works that keep ID features, researchers could investigate ways to introduce ID features in an implicit manner~\cite{li2023ctrl,wang2023flip}, \eg, apply contrastive learning to align between the semantic and collaborative knowledge, and therefore avoid involving ID features for large language models.


\subsection{Fairness}

Researchers have discovered that bias in the pretraining corpus could mislead LLM to generate harmful or offensive content, \eg, discriminating against disadvantaged groups~\cite{sah2024unveiling,deldjoo2024understanding}.
Although there are strategies (\eg, RLHF~\cite{ouyang2022training}) to reduce the harmfulness of LLM, existing works have already detected the unfairness problem in recommender systems brought by LLM from both user-side~\cite{hua2023up5,zhang2023chatgpt} and item-side~\cite{hou2023large} perspectives.

The user-side fairness in recommender systems requires similar users to be treated similarly at either individual level or group level. 
The user sensitive attributes should not be preset during recommendation (\eg, gender, race). 
For instance, \citet{salinas2023unequal} reveal the demographic bias of LLM through job recommendations, where LLM tends to provide unequal opportunities for people with different genders or from different countries.
\citet{xu2023llms} study the traceback, degree, and impact of the implicit user unfairness of LLM for recommendation, and find that LLM will implicitly infer the gender, race or nationality from user name. 
\citet{li2023preliminary} further study to mitigate the provider bias~\cite{burke2018balanced,qi2022profairrec} in news recommendation by either explicitly specifying the number of articles
from both popular and unpopular providers, or explicitly indicating the priority of less popular providers.
To tackle such a user-side unfairness problem, UP5~\cite{hua2023up5} proposes counterfactually fair prompting (CFP), which consists of a personalized prefix prompt and a prompt mixture to ensure fairness w.r.t. a set of sensitive attributes. 
Besides, \citet{zhang2023chatgpt} introduce a benchmark named FaiRLLM, where authors comprise carefully crafted metrics and a dataset that accounts for eight sensitive attributes in recommendation scenarios where LLM is involved. Yet these studies only focus on the fairness issue in specific recommendation tasks (\eg, item generation task) with limited evaluation metrics.

The item-side fairness in recommender systems ensures that each item or item group should receive a fair chance to be recommended (\eg, proportional to its merits or utility)~\cite{patro2020fairrec,liu2019personalized,singh2018fairness}.
However, how to improve item-side fairness in LLM remains less explored.
As a preliminary study, \citet{hou2023large} observe that the popularity bias occurs when LLM serves as a ranking function, and alleviate the bias to some extents by designing prompts to guide the LLM focusing on users' historical interactions. 
Another related work~\cite{liu2023text} alleviates the item popularity bias by representing long-tail items using full-text modeling and bringing the benefits of LLM to recommender systems, but it neglects the intrinsic item-side bias within LLM itself.
Further studies on popularity bias and other potential item-wise fairness issues when adapting LLM to RS are still needed.

\subsection{Other Potential Risks from LLM}

Apart from the fairness problem, researchers have identified many other potential risks that intrinsically stem from large language models, \eg, hallucination~\cite{li2023halueval,wang2023evaluation}. 
When we adapt LLM to RS, some of these biases might be magnified and hurt the reliability of the system.
Hence, in this section, we discuss the new challenges for building harmless and trustworthy LLM-enhanced recommender systems from three perspectives: hallucination, privacy, and explainability. 

\subsubsection{Hallucination}

Hallucination refers to the phenomenon that large language models generate output texts that appear creadible but are actually incorrect or lack of factual basis~\cite{li2023halueval,wang2023evaluation}. 
The hallucination problem of LLM can mislead the recommender system with erroneous information, possibly resulting in recommendation performance degeneration. 
For instance, when adapting LLM to the feature engineering stage of RS for enhancing the item content understanding, a hallucinative output from LLM might erroneously provide fake attributes or descriptions for the given item, adversely affecting the performance of recommendation models. 
Furthermore, the hallucination problem can cause severe risks to individuals, particularly in critical recommendation scenarios like healthcare suggestions, legal guidance and education. 
In these areas, the spread of inaccurate information can lead to serious real repercussions in society. 
Therefore, to counteract the hallucination, it is crucial to verify the correctness and factualness of the generated content from LLM, possibly with the help of external resources like knowledge graphs as the additional verifiable information~\cite{manakul2023selfcheckgpt,mckenna2023sources}.

\subsubsection{Privacy}

The data privacy serves as a long-standing problem in machine learning~\cite{dwork2014algorithmic}, and is becoming increasingly important for recommender systems in the era of large language models due to the following two concerns. 
Firstly, the success of LLM highly relies on the extensive pretraining corpus collected from diverse online sources, some of which might contain users' sensitive information, \eg, the user's email address from social media platforms. 
Secondly, apart from the sensitive information in pretraining corpora, LLM is also frequently leveraged to process or even finetuned on the user behavior data from the recommender system, which encompasses personal preferences, online activities and other identifiable information. 
The accessibility of LLM to these user-sensitive data resources would pose the potential risk of exposing private user information, leading to privacy violations~\cite{carlini2021extracting,yao2023survey}.
Consequently, safeguarding the confidentiality and security of the data is essential for privacy preservation and building a trustworthy recommender system. 
As preliminary studies, DPLLM~\cite{carranza2023privacy} finetunes a differentially private (DP) large language model for privacy-preserved synthetic user query generation in recommender systems. 
\citet{li2023privacy} propose to personalize LLM based on the user's own private data through prompt tuning with a privatized token reconstruction task.

\subsubsection{Explainability}

Generating user-friendly explanations regarding why an item is recommended plays a crucial role in enhancing user trust and facilitating more informed decision makings during recommendation~\cite{luo2023unlocking}.
We discuss the explainability property for LLM-enhanced recommender systems from the following two perspectives.
\textit{Firstly}, LLM can make conventional recommender systems more explainable. 
Several works have revealed that LLM is capable of generating reasonable explanations based on the recommendation output~\cite{ghosh2023jobrecogpt,raczynski2023problem,chu2024llm}, as well as interpreting the latent representations of CRM after careful alignments~\cite{lei2023recexplainer}. 
For instance, \citet{rahdari2023logic} propose the Logic-Scaffolding framework to combine the aspect-based explanation and chain-of-thought prompting for LLM to generate recommendation explanations through intermediate reasoning steps. 
\textit{Secondly}, although LLM helps improve the explainable recommendation, LLM itself is still a black box that lacks explainability for the recommender system, especially when we involve closed-source large language models like ChatGPT and GPT4~\cite{chen2023large}. 
This is potentially risky if the behavior of LLM is unexplainable and uncontrollable when building a reliable and trustworthy LLM-enhanced recommender system.
Based on the two insights above, we argue that the future directions for LLM-enhanced explainable recommendation generally lies in two folds: (1) design better strategies to prompt and acquire recommendation explanations from LLM, and meanwhile (2) seek better ways to enhance the interpretability of LLM itself.

%% file: sections/conclusion.tex
\section{Conclusion and Future Prospects}
\label{sec:conclusion}

In conclusion, large language models have demonstrated impressive human-like capabilities due to their extensive open-world knowledge, the logical and commonsense reasoning ability, and the comprehension of human culture and society~\cite{hadi2023survey,zhao2023survey,wang2023survey}.
As a result, the emergence of large language models is opening up a promising research direction for LLM-enhanced recommender systems. 
This survey proposes a systematic view of the LLM-enhanced recommendation from the perspective of the whole pipeline in industrial recommender systems. 
We comprehensively summarize the latest research progress in adapting large language models to recommender systems from two aspects: where and how to adapt LLM to RS.
\begin{itemize}[leftmargin=10pt]
    \item For the \textbf{``WHERE''} question, we analyze the roles that LLM could play at different stages of the recommendation pipeline, \ie, feature engineering, feature encoder, scoring/ranking function, user interaction, and pipeline controller.
    \item For the \textbf{``HOW''} question, we analyze the training and inference strategies, resulting in two orthogonal classification criteria, \ie, whether to tune LLM during training, and whether to involve CRM for inference.
\end{itemize}

Detailed discussions and insightful development paths are also provided for each taxonomy perspective.
As for future prospects, apart from the three aspects we have already highlighted in Section~\ref{sec: challenge} (\ie, \textbf{efficiency}, \textbf{effectiveness} and \textbf{ethics}), we would like to further express our hopeful vision for the future development of combining large language models and recommender systems:
\begin{itemize}[leftmargin=10pt]
    \item \textbf{A personalized content generation paradigm} for the next-generation recommender systems empowered with AI-Generated Content (AIGC). With the emergence of large language models as well as other multi-modal generative models like diffusion models~\cite{yang2023diffusion,ho2020denoising}, the recommender systems can manage to provide different personalized presentations (\eg, textual content revision, thumbnail generation) of the same item for different users~\cite{wang2023generative}. In this way, we can not only recommend personalized item lists tailored for different users, but also provide different item content presentations of the same item according to diverse user preferences.
    \item \textbf{A unified public benchmark} is of an urgent need to provide reasonable and convincing evaluation protocols, since (1) the fine-grained cross comparison among existing works remains vacant, and (2) it is quite expensive and difficult to reproduce the experimental results of recommendation models combined with LLM. 
    Although there exist some benchmarks for LLM-enhanced RS (\eg, LLMRec~\cite{liu2023llmrec}, OpenP5~\cite{xu2023openp5}), they generally concentrate on a certain aspect of LLM-enhanced RS. 
    For instances, OpenP5~\cite{xu2023openp5} and LLMRec~\cite{liu2023llmrec} only focus on the generative recommendation paradigms that adopt LLM as the scoring/ranking function without help of CRM. 
    Consequently, a unified comparison for the adaptions of LLM to different recommendation pipeline stages (\eg, feature engineering, feature encoder) still remains to be explored.
    \item \textbf{A customized large foundation model} for recommendation domains, which can take over control of the entire recommendation pipeline.
    Currently, research works that involve LLM in the pipeline controller stage generally adopt a frozen general-purpose large foundation model like ChatGPT and GPT4 to connect the different stages. 
    By constructing in-domain instruction data and even customizing the model structure for collaborative knowledge, there is a hopeful vision that we can acquire a large foundation model specially designed for recommendation domains, enabling a new level of automation in recommender systems.
    
\end{itemize}

%% file: sections/appendix.tex
\newpage
\appendix

\section{Look-up Table for Mentioned Works}
\label{sec:look up table}

To provide easy reference for readers and further facilitate the research community on LLM-enhanced recommender systems, we construct a comprehensive lookup table that contains detailed information for works we mentioned in this paper. 
As shown in Table~\ref{tab:look up table}, we first classify the research works according to the stage where their adapted LLM is involved. Different stages are separated and denoted by different colors.
Then, we provide the detailed information for each work (\ie, each row) including but not limited to the size of LLM and the LLM tuning strategy. 
\textbf{Since the research area of LLM-enhanced recommender systems is ever-evolving, readers could refer to our GitHub repository\footnote{\url{https://github.com/CHIANGEL/Awesome-LLM-for-RecSys}} for more up-to-date information about the latest research works.}



{\small\tabcolsep=1pt  
\begin{longtable}{@{}ccccc@{}}
\caption{The look-up table for works on adapting large language models (LLM) to recommender systems (RS) mentioned in this paper. We use the following abbreviations. \textbf{FFT}: full finetuning. \textbf{PT}: prompt tuning. \textbf{LAT}: layerwise adapter tuning. \textbf{OT}: option tuning. \textbf{T-FEW}: few-shot parameter efficient tuning. Note that only the largest models used in the corresponding papers are listed. If the version of the pretrained language model is not specified, we assume it to be the base version. We use \textit{N/A} to denote works that do not name the proposed method.}
\label{tab:look up table}\\
\toprule
\textbf{Model Name} &
  \textbf{LLM Backbone} &
  \textbf{LLM Tuning Strategy} &
  \textbf{RS Task} &
  \textbf{RS Scenario} \\* \midrule
\endfirsthead
\multicolumn{5}{c}%
{{\bfseries Table \thetable\ continued from previous page}} \\
\toprule
\textbf{Model Name} &
  \textbf{LLM Backbone} &
  \textbf{LLM Tuning Strategy} &
  \textbf{RS Task} &
  \textbf{RS Scenario} \\* \midrule
\endhead
\rowcolor[rgb]{0.949,0.504,0.504}\multicolumn{5}{c}{\textbf{Feature Engineering (User- and Item-level Feature  Augmentation)}} \\* \midrule
LLM4KGC \cite{chen2023knowledge} &
  \begin{tabular}[c]{@{}c@{}}PaLM (540B) \\ChatGPT\end{tabular} &
  Frozen &
  N/A &
  E-commerce \\* \midrule
TagGPT \cite{li2023taggpt} &
  ChatGPT &
  Frozen &
  Item Tagging &
  Food, Video \\* \midrule
ICPC \cite{christakopoulou2023large} &
  LaMDA (137B)&
  FFT/PT &
  User Profiling &
  N/A \\* \midrule
KAR \cite{xi2023towards} &
  ChatGPT &
  Frozen &
  \begin{tabular}[c]{@{}c@{}}Reranking\\ CTR Prediction \\ Rating Prediction\end{tabular} &
  N/A \\* \midrule
PIE \cite{brinkmann2023product} &
  ChatGPT &
  Frozen &
  Attribute Extraction &
  E-commerce \\* \midrule
LGIR \cite{du2023enhancing} &
  GhatGLM (6B) &
  Frozen &
  Top-N RS &
  Job \\* \midrule
GIRL \cite{zheng2023generative} &
  BELLE (7B) &
  FFT &
  CTR Prediction &
  Job \\* \midrule
LLM-Rec \cite{lyu2023llm} &
  text-davinci-003 &
  Frozen &
  Top-N RS &
  Movie, Food \\* \midrule
HKFR \cite{yin2023heterogeneous} &
  ChatGPT &
  Frozen &
  Top-N RS &
  E-commerce \\* \midrule
LLaMA-E \cite{shi2023llama} &
  LLaMA (30B) &
  LoRA &
  E-commerce Authoring &
  E-commerce \\* \midrule
EcomGPT \cite{li2023ecomgpt} &
  BLOOMZ (7.1B)	 &
  FFT &
  E-commerce NLP Tasks &
  E-commerce \\* \midrule
TF-DCon \cite{wu2023leveraging} &
  ChatGPT &
  Frozen &
  Dataset Condensation &
  Movie, Book, News \\* \midrule
RLMRec \cite{ren2023representation} &
  ChatGPT &
  Frozen &
  Top-N RS &
  E-commerce, Book, Game \\* \midrule
LLMRec \cite{wei2023llmrec} &
  ChatGPT &
  Frozen &
  Top-N RS &
  Movie, Video \\* \midrule
LLMRG \cite{wang2023enhancing} &
  GPT4 &
  Frozen &
  Top-N RS &
  E-commerce, Movie \\* \midrule
CUP \cite{torbati2023recommendations} &
  ChatGPT &
  Frozen &
  Top-N RS &
  Book \\* \midrule
SINGLE \cite{liu2023modeling} &
  Vicuna (13B) &
  Frozen &
  Sequential RS &
  News \\* \midrule
SAGCN \cite{liu2023understanding} &
  ChatGPT &
  Frozen &
  Top-N RS &
  E-commerce \\* \midrule
UEM \cite{doddapaneni2024user} &
  FLAN-T5-base (250M) &
  FFT &
  User Profiling &
  Movie \\* \midrule
LLMHG \cite{chu2024llm} &
  GPT4 &
  FFT &
  Sequential RS &
  E-commerce, Movie \\* \midrule
Llama4Rec \cite{luo2024integrating} &
  LLaMA2 (7B) &
  FFT &
  \begin{tabular}[c]{@{}c@{}}Top-N RS\\ Sequential RS \\ Rating Prediction\end{tabular} &
  E-commerce, Movie \\* \midrule

\rowcolor[rgb]{0.949,0.504,0.504}\multicolumn{5}{c}{\textbf{Feature Engineering (Instance-level Sample Generation)}} \\* \midrule
GReaT  \cite{borisov2023language} &
  GPT2-Medium (355M) &
  FFT &
  N/A &
  Tabular \\* \midrule
ONCE \cite{liu2023first} &
  ChatGPT &
  Frozen &
  \begin{tabular}[c]{@{}c@{}}Retrieval\\ Sequential RS\end{tabular} &
  News \\* \midrule
AnyPredict \cite{wang2023anypredict} &
  ChatGPT &
  Frozen &
  N/A &
  Tabular \\* \midrule
DPLLM \cite{carranza2023privacy} &
  T5-XL (3B) &
  FFT &
  \begin{tabular}[c]{@{}c@{}}Retrieval\\Privacy\end{tabular} &
  Web Search \\* \midrule
MINT \cite{mysore2023large} &
  text-davinci-003 &
  Frozen &
  Narrative-Driven RS &
  POI \\* \midrule
Agent4Rec \cite{zhang2023generative} &
  ChatGPT &
  Frozen &
  RS Simulation &
  Movie \\* \midrule
RecPrompt \cite{liu2023recprompt} &
  GPT4 &
  Frozen &
  Top-N RS &
  News \\* \midrule
PO4ISR \cite{sun2023large} &
  GPT4 &
  Frozen &
  Session-based RS &
  E-commerce, Movie, Game \\* \midrule
BEQUE \cite{li2023agent4ranking} &
  ChatGLM (6B) &
  FFT &
  Query Rewriting &
  E-commerce \\* \midrule
Agent4Ranking \cite{li2023agent4ranking} &
  GPT4 &
  Frozen &
  Query Rewriting &
  Web Search \\* \midrule

\rowcolor[rgb]{0.582,0.879,0.824}\multicolumn{5}{c}{\textbf{Feature Encoder (Representation Enhancement)}} \\* \midrule
U-BERT \cite{qiu2021u} &
  BERT-base (110M) &
  FFT &
  Rating Prediction &
  \begin{tabular}[c]{@{}c@{}}Business\\ E-commerce\end{tabular} \\* \midrule
UNBERT \cite{zhang2021unbert} &
  BERT-base (110M) &
  FFT &
  Sequential RS &
  News \\* \midrule
PLM-NR \cite{wu2021empowering} &
  RoBERTa-base (125M) &
  FFT &
  Sequential RS &
  News \\* \midrule
Pyramid-ERNIE \cite{zou2021pre} &
  ERNIE (110M) &
  FFT &
  Ranking &
  Web Search \\* \midrule
ERNIE-RS \cite{liu2021pre} &
  ERNIE (110M) &
  FFT &
  Retrieval &
  Web Search \\* \midrule
CTR-BERT \cite{muhamed2021ctr} &
  Customized BERT (1.5B) &
  FFT &
  CTR Prediction &
  E-commerce \\* \midrule
SuKD \cite{wang2022learning} &
  RoBERTa-large (355M) &
  FFT &
  CTR Prediction &
  Advertisement \\* \midrule
PREC \cite{liu2022boosting} &
  BERT-base (110M) &
  FFT &
  CTR Prediction &
  News \\* \midrule
MM-Rec \cite{wu2022mm} &
  BERT-base (110M) &
  FFT &
  Sequential RS &
  News \\* \midrule
Tiny-NewsRec \cite{yu2022tiny} &
  UniLMv2-base (110M) &
  FFT &
  Sequential RS &
  News \\* \midrule
PTM4Tag \cite{he2022ptm4tag} &
  CodeBERT (125M) &
  FFT &
  Top-N RS &
  posts \\* \midrule
TwHIN-BERT \cite{zhang2022twhin} &
  BERT-base (110M) &
  FFT &
  Social RS &
  posts \\* \midrule
LSH \cite{rahmani2023improving} &
  BERT-base (110M) &
  FFT &
  Top-N RS &
  Code \\* \midrule
LLM2BERT4Rec \cite{harte2023leveraging} &
  text-embedding-ada-002 &
  Frozen &
  Sequential RS &
  E-commerce \\* \midrule
LLM4ARec \cite{li2023prompt} &
  GPT2 (110M) &
  PT &
  Aspect-based RS &
  E-commerce, Movie \\* \midrule
TIGER \cite{rajput2023recommender} &
  Sentence-T5-base (223M) &
  Frozen &
  Sequential RS &
  E-commerce \\* \midrule
TBIN \cite{chen2023tbin} &
  BERT-base (110M) &
  Frozen &
  CTR Prediction &
  E-commerce \\* \midrule
LKPNR \cite{runfeng2023lkpnr} &
  LLaMA2 (7B) &
  Frozen &
  Sequential RS &
  News \\* \midrule
SSNA \cite{peng2023towards} &
  DistilRoBERTa-base (83M) &
  LAT &
  Sequential RS &
  E-commerce \\* \midrule
CollabContext \cite{wang2023collaborative} &
  Instructor-XL (1.5B) &
  Frozen &
  Top-N RS &
  E-commerce \\* \midrule
LMIndexer \cite{jin2023language} &
  T5-base (223M) &
  FFT &
  \begin{tabular}[c]{@{}c@{}}Sequential RS \\ Product Search \\Document Retrieval\end{tabular} &
  E-commerce \\* \midrule
Stack \cite{putatunda2023bert} &
  BERT-base (110M) &
  Frozen &
  Top-N RS &
  E-commerce \\* \midrule
N/A \cite{ho2023utilizing} &
  BERT-base (110M) &
  FFT &
  Sequential RS &
  POI \\* \midrule
UEM \cite{doddapaneni2024user} &
  Sentence-T5-base (223M) &
  Frozen &
  User Profiling &
  Movie \\* \midrule
Social-LLM \cite{jiang2023social} &
  SBERT-MPNet-base (110M) &
  Frozen &
  Social RS &
  Social Network \\* \midrule
LLMRS \cite{john2024llmrs} &
  MPNet (110M) &
  Frozen &
  Sequential RS &
  E-commerce \\* \midrule

\rowcolor[rgb]{0.582,0.879,0.824}\multicolumn{5}{c}{\textbf{Feature Encoder (Unified Cross-domain Recommendation)}} \\* \midrule
ZESRec \cite{ding2021zero} &
  BERT-base (110M) &
  Frozen &
  Sequential RS &
  E-commerce \\* \midrule
UniSRec \cite{hou2022towards} &
  BERT-base (110M) &
  Frozen &
  Sequential RS &
  E-commerce \\* \midrule
VQ-Rec \cite{hou2023learning} &
  BERT-base (110M) &
  Frozen &
  Sequential RS &
  E-commerce \\* \midrule
IDRec  vs MoRec \cite{yuan2023go} &
  BERT-base (110M) &
  FFT &
  Sequential RS &
  E-commerce, News, Video \\* \midrule
TransRec \cite{fu2023exploring} &
  RoBERTa-base (125M) &
  LAT &
  \begin{tabular}[c]{@{}c@{}}Cross-domain RS\\ Sequential RS\end{tabular} &
  E-commerce, News, Video \\* \midrule
TCF \cite{li2023tcf} &
  OPT-175B (175B) &
  Frozen/FFT &
  Top-N RS &
  Fashion, News, Video \\* \midrule
S\&R Foundation \cite{gong2023unified} &
  ChatGLM (6B) &
  Frozen &
  \begin{tabular}[c]{@{}c@{}}CTR Prediction \\ Ranking \\ Relevance Prediction\end{tabular} &
  E-commerce \\* \midrule
MISSRec \cite{wang2023missrec} &
  CLIP-B/32 (400M) &
  FFT &
  Sequential RS &
  E-commerce \\* \midrule
UFIN \cite{tian2023ufin} &
  FLAN-T5-base (250M) &
  Frozen &
  CTR Prediction &
  E-commerce, Movie \\* \midrule
PMMRec \cite{li2023multi} &
  RoBERTa-large (355M) &
  Top-2-layer FT &
  Multi-modal RS &
  E-commerce, Video \\* \midrule
Uni-CTR \cite{fu2023unified} &
  Sheared-LLaMA (1.3B) &
  LoRA &
  CTR Prediction &
  E-commerce \\* \midrule

\rowcolor[rgb]{0.836,0.965,0.676}\multicolumn{5}{c}{\textbf{Scoring/Ranking Function (Item Scoring Task)}} \\* \midrule
LMRecSys \cite{zhang2021language} &
  GPT2-XL (1.5B) &
  FFT &
  Top-N RS &
  Movie \\* \midrule
PTab \cite{liu2022ptab} &
  BERT-base (110M) &
  FFT &
  N/A &
  Tabular \\* \midrule
UniTRec \cite{mao2023unitrec} &
  BART (406M) &
  FFT &
  Sequential RS &
  News, Social Media \\* \midrule
Prompt4NR \cite{zhang2023prompt} &
  BERT-base/110M &
  FFT &
  Sequential RS &
  News \\* \midrule
RecFormer \cite{li2023text} &
  LongFormer/149M &
  FFT &
  Sequential RS &
  Product \\* \midrule
TabLLM \cite{hegselmann2023tabllm} &
  T0 (11B) &
  T-FEW &
  N/A &
  Tabular \\* \midrule
Zero-shot GPT \cite{sileo2022zero} &
  GPT2-Medium (355M) &
  Frozen &
  Rating Prediction &
  Movie \\* \midrule
FLAN-T5 \cite{kang2023llms} &
  FLAN-T5-XXL (11B) &
  FFT &
  Rating Prediction &
  Book, Movie \\* \midrule
BookGPT \cite{zhiyuli2023bookgpt} &
  ChatGPT &
  Frozen &
  \begin{tabular}[c]{@{}c@{}}Sequential RS\\ Top-N RS\\ Summary Recommendation\end{tabular} &
  Book \\* \midrule
TALLRec \cite{bao2023tallrec} &
  LLaMA (7B) &
  LoRA &
  Sequential RS &
  Book, Movie \\* \midrule
PBNR \cite{li2023pbnr} &
  T5-small (60M) &
  FFT &
  Sequential RS &
  News \\* \midrule
CR-SoRec \cite{prakash2023cr} &
  BERT-base (110M) &
  FFT &
  Social RS &
  Social Media, E-commerce \\* \midrule
PromptRec \cite{wu2023towards} &
  LLaMA (7B) &
  Frozen &
  CTR Prediction &
  E-commerce, Movie \\* \midrule
GLRec \cite{wu2023exploring} &
  BELLE-LLaMA (7B) &
  LoRA &
  Top-N RS &
  Job \\* \midrule
BERT4CTR \cite{wang2023bert4ctr} &
  RoBERTa-large (355M) &
  FFT &
  CTR Prediction &
  Advertisement \\* \midrule
ReLLa \cite{lin2023rella} &
  Vicuna (13B) &
  LoRA &
  Sequential RS &
  Movie, Book \\* \midrule
TASTE \cite{liu2023text} &
  T5-base (223M) &
  FFT &
  Sequential RS &
  E-commerce \\* \midrule
N/A \cite{torbati2023unveiling} &
  BERT-base (110M) &
  FFT &
  Top-N RS &
  Book \\* \midrule
ClickPrompt \cite{lin2023clickprompt} &
  RoBERTa-large (355M) &
  FFT &
  CTR Prediction &
  E-commerce, Movie, Book \\* \midrule
SetwiseRank \cite{zhuang2023setwise} &
  FLAN-T5-XXL (11B) &
  Frozen &
  Ranking &
  Web Search \\* \midrule
UPSR \cite{qu2023thoroughly} &
  T5-base (223M) &
  FFT &
  Sequential RS &
  E-commerce \\* \midrule
LLM-Rec \cite{tang2023one} &
  OPT (6.7B) &
  LoRA &
  Sequential RS &
  E-commerce \\* \midrule
LLMRanker \cite{zhuang2023beyond} &
  FLAN PaLM2 S &
  Frozen &
  Ranking &
  Web Search\\* \midrule
CoLLM \cite{zhang2023collm} &
  Vicuna (7B) &
  LoRA &
  CTR Prediction &
  Movie, Book \\* \midrule
FLIP \cite{wang2023flip} &
  RoBERTa-large (355M) &
  FFT &
  CTR Prediction &
  Movie, Book \\* \midrule
BTRec \cite{ho2023btrec} &
  BERT-base (110M) &
  FFT &
  Sequential RS &
  POI \\* \midrule
CLLM4Rec \cite{zhu2023collaborative} &
  GPT2 (110M) &
  FFT &
  Sequential RS &
  E-commerce \\* \midrule
CUP \cite{torbati2023recommendations} &
  BERT-base (110M) &
  Last-layer FT &
  Top-N RS &
  Book \\* \midrule
N/A \cite{sun2023instruction} &
  FLAN-T5-XL (3B) &
  FFT &
  Ranking &
  Web Search \\* \midrule
CoWPiRec \cite{yang2023collaborative} &
  BERT-base (110M) &
  FFT &
  Sequential RS &
  E-commerce \\* \midrule
RecExplainer \cite{lei2023recexplainer} &
  Vicuna-v1.3 (7B) &
  LoRA &
  Sequential RS &
  E-commerce \\* \midrule
E4SRec \cite{li2023e4srec} &
  LLaMA2 (13B) &
  LoRA &
  Sequential RS &
  E-commerce \\* \midrule
CER \cite{raczynski2023problem} &
  GPT2 (110M)	 &
  FFT &
  Explainable RS &
  E-commerce, Movie \\* \midrule
LSAT \cite{shi2023preliminary} &
  LLaMA (7B) &
  LoRA &
  Sequential RS &
  E-commerce, Movie \\* \midrule
Llama4Rec \cite{luo2024integrating} &
  LLaMA2 (7B) &
  FFT &
  \begin{tabular}[c]{@{}c@{}}Top-N RS\\ Sequential RS \\ Rating Prediction\end{tabular} &
  Movie, Book \\* \midrule

\rowcolor[rgb]{0.836,0.965,0.676}\multicolumn{5}{c}{\textbf{Scoring/Ranking Function (Item Generation Task)}} \\* \midrule
GPT4Rec \cite{li2023gpt4rec} &
  GPT2 (110M) &
  FFT &
  Sequential RS &
  E-commerce \\* \midrule
UP5 \cite{hua2023up5} &
  T5-base (223M) &
  FFT &
  \begin{tabular}[c]{@{}c@{}}Retrieval\\ Sequential RS\end{tabular} &
  Movie, Insurance \\* \midrule
VIP5 \cite{geng2023vip5} &
  T5-base (223M) &
  LAT &
  \begin{tabular}[c]{@{}c@{}}Sequential RS\\ Top-N RS\\ Explanation Generation\end{tabular} &
  E-commerce \\* \midrule
P5-ID \cite{hua2023index} &
  T5-small (61M) &
  FFT &
  Sequential RS &
  \begin{tabular}[c]{@{}c@{}}Business\\ E-commerce\end{tabular} \\* \midrule
FaiRLLM \cite{zhang2023chatgpt} &
  ChatGPT &
  Frozen &
  Top-N RS &
  Music, Movie \\* \midrule
PALR \cite{chen2023palr} &
  LLaMA (7B) &
  FFT &
  Sequential RS &
  E-commerce, Movie \\* \midrule
ChatGPT-3 \cite{hou2023large} &
  ChatGPT &
  Frozen &
  Sequential RS &
  E-commerce, Movie \\* \midrule
AGR \cite{lin2023sparks} &
  ChatGPT &
  Frozen &
  Conversational RS &
  N/A \\* \midrule
NIR \cite{wang2023zero} &
  GPT-3 (175B) &
  Frozen &
  Sequential RS &
  Movie \\* \midrule
GPTRec \cite{petrov2023generative} &
  GPT2-medium (355M) &
  FFT &
  Top-N RS &
  E-commerce, Movie, Music \\* \midrule
ChatNews \cite{li2023preliminary} &
  ChatGPT &
  Frozen &
  Sequential RS &
  News \\* \midrule
N/A \cite{sanner2023large} &
  PaLM (62B) &
  Frozen &
  Sequential RS &
  Movie \\* \midrule
LLMSeqPrompt \cite{harte2023leveraging} &
  OpenAI ada model &
  FT &
  Sequential RS &
  E-commerce \\* \midrule
GenRec \cite{ji2023genrec} &
  LLaMA (7B) &
  LoRA &
  Sequential RS &
  E-commerce, Movie \\* \midrule
HKFR \cite{yin2023heterogeneous} &
  ChatGLM (6B) &
  LoRA &
  Top-N RS &
  POI \\* \midrule
N/A \cite{salinas2023unequal} &
  ChatGPT &
  Frozen &
  Fair RS &
  Job \\* \midrule
BIGRec \cite{bao2023bi} &
  LLaMA (7B) &
  LoRA &
  Sequential RS &
  Movie, Game \\* \midrule
KP4SR \cite{zhai2023knowledge} &
  T5-small (60M) &
  FFT &
  Sequential RS &
  Movie, Music, Book \\* \midrule
RecSysLLM \cite{chu2023leveraging} &
  GLM (10B) &
  LoRA &
  \begin{tabular}[c]{@{}c@{}} Top-N RS\\ Sequential RS\\ Explanation Generation\\ Review Summarization\end{tabular} &
  E-commerce \\* \midrule
POD \cite{li2023distillation} &
  T5-small (60M)	 &
  FFT &
  \begin{tabular}[c]{@{}c@{}} Top-N RS\\ Sequential RS\\ Explanation Generation\end{tabular} &
  E-commerce \\* \midrule
N/A \cite{di2023evaluating} &
  ChatGPT	 &
  Frozen &
  \begin{tabular}[c]{@{}c@{}}Reranking\\ Top-N RS\end{tabular} &
  Movie, Music, Book \\* \midrule
RaRS \cite{di2023retrieval} &
  ChatGPT	 &
  Frozen &
  Top-N RS &
  Movie, Book \\* \midrule
JobRecoGPT \cite{ghosh2023jobrecogpt} &
  GPT4	 &
  Frozen &
  Top-N RS &
  Job \\* \midrule
LANCER \cite{jiang2023reformulating} &
  GPT2 (110M) &
  Prefix Tuning &
  Sequential RS &
  Movie, Books, News \\* \midrule
TransRec \cite{lin2023multi} &
  LLaMA (7B)	 &
  LoRA &
  Sequential RS &
  E-commerce \\* \midrule
AgentCF \cite{zhang2023agentcf} &
  \begin{tabular}[c]{@{}c@{}} text-davinci-003\\ gpt-3.5-turbo\end{tabular} &
  Frozen &
  Sequential RS &
  E-commerce \\* \midrule
P4LM \cite{jeong2023factual} &
  PaLM2-XS &
  FFT &
  Top-N RS &
  Movie \\* \midrule
InstructMK \cite{wang2023multiple} &
  LLaMA (7B) &
  FFT &
  Top-N RS &
  Movie \\* \midrule
LightLM \cite{mei2023lightlm} &
  T5-small (60M) &
  FFT &
  Top-N RS &
  E-commerce \\* \midrule
LlamaRec \cite{yue2023llamarec} &
  LLaMA2 (7B)	 &
  QLoRA &
  Sequential RS &
  E-commerce, Movie, Game \\* \midrule
N/A \cite{zhou2023exploring} &
  GPT-4V	 &
  Frozen &
  Multi-modal RS &
  \begin{tabular}[c]{@{}c@{}} Culture, Art, Media\\ Entertainment, Retail\end{tabular} \\* \midrule
N/A \cite{li2023exploring} &
  ChatGPT &
  gpt-3.5-turbo FT API &
  Top-N RS &
  News \\* \midrule
N/A \cite{xu2023llms} &
  ChatGPT &
  Frozen &
  Fair RS &
  News, Job \\* \midrule
LC-Rec \cite{zheng2023adapting} &
  LLaMA (7B) &
  LoRA &
  Sequential RS &
  E-commerce \\* \midrule
DOKE \cite{zheng2023adapting} &
  ChatGPT &
  Frozen &
  Top-N RS &
  E-commerce, Movie \\* \midrule
ControlRec \cite{qiu2023controlrec} &
  T5-base (223M)	 &
  FFT &
  Sequential RS &
  E-commerce \\* \midrule
LLaRA \cite{liao2023llara} &
  LLaMA2 (7B) &
  LoRA &
  Sequential RS &
  Movie, Game \\* \midrule
PO4ISR \cite{sun2023large} &
  ChatGPT &
  Frozen &
  Session-based RS &
  E-commerce, Movie, Game \\* \midrule
DRDT \cite{wang2023drdt} &
  ChatGPT &
  Frozen &
  Sequential RS &
  E-commerce, Movie \\* \midrule
RecPrompt \cite{liu2023recprompt} &
  GPT4 &
  Frozen &
  Top-N RS &
  News \\* \midrule
LiT5 \cite{singh2023scaling} &
  T5-XL (3B) &
  FFT &
  Ranking &
  Web Search \\* \midrule
STELLA \cite{ma2023large} &
  ChatGPT &
  Frozen &
  Sequential RS &
  Movie, Book, Music, News \\* \midrule
Llama4Rec \cite{luo2024integrating} &
  LLaMA2 (7B) &
  FFT &
  \begin{tabular}[c]{@{}c@{}}Top-N RS\\ Sequential RS \\ Rating Prediction\end{tabular} &
  E-commerce, Movie \\* \midrule

\rowcolor[rgb]{0.836,0.965,0.676} \multicolumn{5}{c}{\textbf{Scoring/Ranking Function (Hybrid Task)}} \\* \midrule
P5 \cite{geng2022recommendation} &
  T5-base (223M) &
  FFT &
  \begin{tabular}[c]{@{}c@{}}Rating Prediction\\ Top-N RS\\ Sequential RS\\ Explanation Generation\\ Review Summarization\end{tabular} &
  E-commerce, Business \\* \midrule
M6-Rec \cite{cui2022m6} &
  M6-base (300M) &
  OT &
  \begin{tabular}[c]{@{}c@{}}Retrieval\\ Ranking\\ Explanation Generation\\ Conversational RS\end{tabular} &
  E-commerce \\* \midrule
InstructRec \cite{zhang2023recommendation} &
  Flan-T5-XL (3B) &
  FFT &
  \begin{tabular}[c]{@{}c@{}}Sequential RS\\ Product Search\\ Personalized Search\\ Matching-then-reranking\end{tabular} &
  E-commerce \\* \midrule
ChatGPT-1 \cite{liu2023chatgpt} &
  ChatGPT &
  Frozen &
  \begin{tabular}[c]{@{}c@{}}Rating Prediction\\ Top-N RS\\ Sequential RS\\ Explanation Generation\\ Review Summarization\end{tabular} &
  E-commerce \\* \midrule
ChatGPT-2 \cite{dai2023uncovering} &
  ChatGPT &
  Frozen &
  \begin{tabular}[c]{@{}c@{}}Pointwise Scoring\\ Pairwise Comparison\\ Listwise Ranking\end{tabular} &
  E-commerce, Movie, News \\* \midrule
ChatGPT-4 \cite{sun2023chatgpt} &
  ChatGPT &
  Frozen &
  Passage Reranking &
  Web Search \\* \midrule
BDLM \cite{zhang2023bridging} &
  Vicuna (7B) &
  FFT &
  Top-N RS &
  E-commerce, Movie, Luxury \\* \midrule
RecRanker \cite{luo2023recranker} &
  LLaMA2 (13B) &
  FFT &
  \begin{tabular}[c]{@{}c@{}}Pointwise Scoring\\ Pairwise Comparison\\ Listwise Ranking\end{tabular} &
  Movie, Book \\* \midrule

\rowcolor{UserInter}\multicolumn{5}{c}{\textbf{User Interaction (Task-oriented User Interaction)}} \\* \midrule
TG-ReDial	 \cite{zhou2020tgredial} &
  \begin{tabular}[c]{@{}c@{}}BERT-base (110M) \\ GPT2 (110M)\end{tabular} &
  Unknown &
  Conversational RS &
  Movie \\* \midrule
TCP \cite{wang2022follow} &
  BERT-base (110M) &
  FFT &
  Conversational RS &
  \begin{tabular}[c]{@{}c@{}}Movie, Music, Food \\ Restaurant, News, Weather\end{tabular} \\* \midrule
MESE \cite{yang2022mese} &
  \begin{tabular}[c]{@{}c@{}}DistilBERT (67M) \\ GPT2 (110M)\end{tabular} &
  FFT &
  Conversational RS &
  Movie \\* \midrule
UniMIND \cite{deng2023unimind} &
  BART-base (139M) &
  FFT &
  Conversational RS &
  \begin{tabular}[c]{@{}c@{}}Movie, Music, Food \\ Restaurant, News, Weather\end{tabular} \\* \midrule
VRICR \cite{zhang2023vricr} &
  BERT-base (110M) &
  FFT &
  Conversational RS &
  Movie \\* \midrule
KECR \cite{ren2023kecr} &
  \begin{tabular}[c]{@{}c@{}}BERT-base (110M) \\ GPT2 (110M)\end{tabular} &
  Frozen &
  Conversational RS &
  Movie \\* \midrule
N/A \cite{he2023large} &
  GPT4 &
  Frozen &
  Conversational RS &
  Movie \\* \midrule
MuseChat \cite{dong2023musechat} &
  Vicuna (7B)	 &
  LoRA &
  Conversational RS &
  Music, Video \\* \midrule
N/A \cite{liu2023conversational} &
  Chinese-Alpaca (7B) &
  LoRA &
  Conversational RS &
  E-commerce \\* \midrule

\rowcolor{UserInter}\multicolumn{5}{c}{\textbf{User Interaction (Open-ended User Interaction)}} \\* \midrule
BARCOR \cite{wang2022barcor} &
  BART-base (139M) &
  Selective-layer FT	 &
  Conversational RS &
  Movie \\* \midrule
RecInDial \cite{wang2022recindial} &
  DialoGPT (110M) &
  FFT	 &
  Conversational RS &
  Movie \\* \midrule
UniCRS \cite{wang2022unicrs} &
  DialoGPT-small (176M) &
  Frozen	 &
  Conversational RS &
  Movie \\* \midrule
T5-CR \cite{ram2023multitask} &
  T5-base (223M) &
  FFT	 &
  Conversational RS &
  Movie \\* \midrule
TtW \cite{leszczynski2023talktw} &
  \begin{tabular}[c]{@{}c@{}}T5-base (223M) \\ T5-XXL (11B)\end{tabular} &
  \begin{tabular}[c]{@{}c@{}}FFT \\ Frozen\end{tabular} &
  Conversational RS &
  Music \\* \midrule
T5-CR \cite{wang2023rethinking} &
  ChatGPT &
  Frozen	 &
  Conversational RS &
  Movie, Books, Sports, Music \\* \midrule

\rowcolor{gray}\multicolumn{5}{c}{\textbf{Pipeline Controller}} \\* \midrule
Chat-REC \cite{gao2023chat} &
  ChatGPT &
  Frozen &
  \begin{tabular}[c]{@{}c@{}} Rating Prediction\\Top-N RS \end{tabular}&
  Movie \\* \midrule
RecLLM \cite{friedman2023leveraging} &
  LLaMA (7B) &
  FFT &
  Conversational RS &
  Video \\* \bottomrule
RAH \cite{shu2023rah} &
  GPT4 &
  Frozen &
  Top-N RS &
  Movie, Book, Game \\* \bottomrule
RecMind \cite{wang2023recmind} &
  ChatGPT &
  Frozen &
  \begin{tabular}[c]{@{}c@{}}Rating Prediction\\ Top-N RS\\ Sequential RS\\ Explanation Generation\\ Review Summarization\end{tabular} &
  Beauty, Business \\* \midrule
InteRecAgent \cite{huang2023recommender} &
  GPT4 &
  Frozen &
  Conversational RS &
  E-commerce, Movie, Game \\* \midrule
CORE \cite{jin2023lending} &
  N/A &
  N/A &
  Conversational RS &
  E-commerce, Movie, Music, Book \\* \midrule
\end{longtable}
}

